\documentclass[aps,prd,onecolumn,11pt,floatfix,nofootinbib,superscriptaddress,preprintnumbers]{revtex4-2}
\usepackage{selinput}
\usepackage[T1]{fontenc}
\usepackage{graphicx}
\usepackage{epstopdf}
\usepackage{amsmath,amssymb}
\usepackage{verbatim}
\usepackage[dvipsnames]{xcolor}
\usepackage{units}
\usepackage[normalem]{ulem}
\usepackage{systeme}
\usepackage{amsmath}
\usepackage[toc,page]{appendix}
\usepackage{float}
\usepackage{lipsum}
\usepackage{hyperref}
\usepackage{caption}
\newcommand{\kbu}{{k_{\rm bu}}}
\newcommand{\ksh}{{k_{\rm sh}}}
\newcommand{\qbu}{{q_{\rm bu}}}

\newcommand{\Nc}{N_{\rm c}}

\newcommand{\fFD}{f_{\rm FD}}
\newcommand{\fB}{f_{\rm B}}
\newcommand{\fQ}{f_{\rm Q}}
\newcommand{\betaid}{\hat{\beta}}%{\beta_{\rm id}}
\newcommand{\muid}{\hat{\mu}}%{\mu_{\rm id}}
\newcommand{\Tid}{\hat{T}}%{T_{\rm id}}
\newcommand{\prob}{\mathcal{P}}
\newcommand{\muB}{\mu_{B}}

\newcommand{\bq}{\boldsymbol{q}}
\newcommand{\bk}{\boldsymbol{k}}
\newcommand{\bE}{\boldsymbol{E}}
\newcommand{\bn}{\boldsymbol{n}}
\newcommand{\bN}{\boldsymbol{N}}
\newcommand{\bg}{\boldsymbol{g}}
\newcommand{\kF}{k_{\mathrm{F}}}
\newcommand{\nB}{n_{\mathrm{B}}}
\newcommand{\eB}{\varepsilon_{\mathrm{B}}}
\newcommand{\eQ}{\varepsilon_{\mathrm{Q}}}

\begin{document}
\title{Statistical Mechanics of Quarkyonic Matter}

\author{Marcus Bluhm}
\email{bluhm@subatech.in2p3.fr}
\affiliation{SUBATECH UMR 6457 (IMT Atlantique, Universit\'e de Nantes, IN2P3/CNRS)\\ 4 rue Alfred Kastler, 44307 Nantes, France}
\author{Yuki Fujimoto}
\affiliation{Department of Physics, Niigata University, Ikarashi, Niigata 950-2181, Japan}
\affiliation{RIKEN Center for Interdisciplinary Theoretical and Mathematical Sciences (iTHEMS), RIKEN, Wako 351-0198, Japan}
\author{Marlene Nahrgang}
\affiliation{SUBATECH UMR 6457 (IMT Atlantique, Universit\'e de Nantes, IN2P3/CNRS)\\ 4 rue Alfred Kastler, 44307 Nantes, France}

\date{\today}

\begin{abstract}
We extend the theoretical formulation of Quarkyonic Matter within the IdylliQ model framework proposed in [Y.~Fujimoto et al., Phys.\ Rev.\ Lett.\ 132, 112701 (2024)~\cite{Fujimoto:2023mzy}] for zero temperature to non-zero temperatures. To this end, we develop a consistent statistical mechanics and grand canonical ensemble description of Quarkyonic Matter as a quantum system subject to additional inequality constraints due to the Pauli exclusion principle acting simultaneously on baryons and their constituent quarks. These constraints result in a significant reduction in the number of physically available baryon states compared to an ideal Fermi gas. As a consequence, the one-particle baryon distribution function factorizes into a thermal Fermi-Dirac distribution and a momentum-dependent density of states. This separation allows us to derive a proper definition of the entropy density that satisfies the third law of thermodynamics in the zero-temperature limit. Moreover, we find that inside Quarkyonic Matter the physical temperature and the physical baryon chemical potential differ from the Lagrange multipliers appearing in the Fermi-Dirac distribution which may have important consequences for the thermodynamics of Quarkyonic Matter.
\end{abstract}

\maketitle
\tableofcontents

\section{Introduction}

An interesting phase of nuclear matter, known as Quarkyonic Matter, has been proposed to exist at high baryon densities~\cite{McLerran:2007qj}. This idea was originally  developed in the large-$N_c$ limit, where $N_c$ is the number of quark colors. At zero temperature, this phase is expected to emerge at a baryon chemical potential, $\muB$, that lies between the characteristic QCD scale, $\muB \sim M_N$, and the deconfinement scale~\cite{LeBellac:1996}, $\muB \sim \sqrt{N_c} M_N$, where $M_N$ is the nucleon mass (see also discussion in Refs.~\cite{Bluhm:2024uhj, Fujimoto:2025sxx}). 
In this peculiar state of matter, quarks are confined inside baryonic states but their characteristics impact the properties of the matter. An important aspect is that the quark degrees of freedom saturate and their occupation numbers are limited by the fermionic spin statistics.
Subsequent work~\cite{McLerran:2018hbz, Jeong:2019lhv, Duarte:2020xsp, *Duarte:2020kvi, Zhao:2020dvu, Margueron:2021dtx, Fujimoto:2023mzy, Moss:2024uam, Saito:2025yld} argued that even for $N_c = 3$ colors Quarkyonic Matter exhibits properties that naturally lead to a very rapid stiffening of the equation of state, potentially explaining recent neutron star observations (see, e.g., Refs.~\cite{Bedaque:2014sqa, Tews:2018kmu, Fujimoto:2019hxv, *Fujimoto:2021zas, *Fujimoto:2024cyv, Drischler:2020fvz}).
This arises due to the specific phase-space structure of baryons and quarks, which in the dual description of the IdylliQ model~\cite{Fujimoto:2023mzy, Kojo:2021ugu} are related via
\begin{equation}
  \label{eq:duality}
  \fQ (q)
  = \int \!\! \frac{d^3 \bk}{(2\pi)^3} \, \varphi\left(\bq - \frac{\bk}{N_c}\right) \fB(k) 
\end{equation}
for the baryon distribution function $\fB(k)$ and the quark distribution function $\fQ(q)$ for a fixed color,~i.e. $\fQ \equiv \fQ^{R} = \fQ^{G} = \fQ^{B} $. The kernel function $\varphi$ describes the momentum distribution of a single quark in a single baryon state, effectively taking into account the confinement of quarks inside baryons.
In the IdylliQ model~\cite{Fujimoto:2023mzy}, we choose the following form for $\varphi$,
\begin{equation}
  \label{eq:wf3d}
  \varphi(\bq) = \frac{2\pi^2}{\Lambda^3} \frac{e^{-q/\Lambda}}{q/\Lambda}\,,
\end{equation}
where the parameter $\Lambda$ is an effective confinement scale. The kernel $\varphi$ is also the Green's function of the differential operator 
\begin{equation}
\hat{L}_q \equiv -\nabla_q^2 + 1/\Lambda^2\, ,
\label{eq:diffOpL}
\end{equation}
by the virtue of which we can invert the duality relation in Eq.~(\ref{eq:duality}) to get the correspondance $\fB(\Nc q) = (\Lambda^2 / \Nc^3) \,\hat{L}_q [\fQ(q)]$. With the explicit form in Eq.~\eqref{eq:wf3d}, the model becomes analytically solvable. For Quarkyonic Matter, the quark degrees of freedom saturate for some low quark-momentum region $q\leq\qbu$ as the Pauli exclusion principle forces the quark distribution function to saturate, $\fQ(q\leq\qbu)=1$. As a direct consequence one finds $\fB(k\leq\kbu)=1/\Nc^3$, where $k=N_c\,q$. The low-momentum part of the quark and baryon distributions is thus completely determined by duality and does not explicitly depend on the thermodynamics up to $\kbu=N_c\,\qbu$ (which depends itself on the medium, as we will see later). The bulk part of the distribution is thus locked by an additional constraint, the Pauli exclusion principle applied to the quarks.

At zero temperature, $T=0$, the physical solution for the entire momentum range of $\fB$ and $\fQ$ can be found via the variational principle of minimizing the energy density $\varepsilon$ for a given baryon density $\nB$ while simultaneously satisfying the fermionic occupation constraints for the baryonic degrees of freedom $0 \leq \fB \leq 1$ and for the quark degrees of freedom $0 \leq \fQ \leq 1$~\cite{Fujimoto:2023mzy}. In the normal nuclear gas phase, one has $\kbu = 0$ and $\fB(k) = \Theta(\kF - k)$, where the energy at the Fermi momentum $k_F$ is identical to the baryon chemical potential, $E(k_F) = \mu_B$. Then we can write the zero temperature limit of the baryon distribution also as $\fB(E) = \Theta(\mu_B-E)$. As the baryon density increases, normal nuclear matter turns into Quarkyonic Matter once the associated quark distribution function at vanishing quark momentum saturates the Pauli exclusion principle constraint, $\fQ(q=0)=1$. This marks the onset of Quarkyonic Matter~\cite{Bluhm:2024uhj}. Beyond the onset point for even larger $\nB$, the baryon distribution function is described by a shell structure as~\cite{Fujimoto:2023mzy}
\begin{equation}
    \fB(k) = \frac{1}{N_c^3}\Theta(\kbu - k) + \Theta(\ksh - k) \Theta(k - \kbu)\, .
    \label{eq:fB}
\end{equation}
The bulk part at low baryon-momentum up to $\kbu$ shows a reduction by a factor of $1/N_c^3$. In this region $\fQ(q \leq \kbu / N_c)=1$. For momenta in the shell $\kbu < k < \ksh$, the baryon distribution is unity, and for momenta larger than $\ksh$ the baryon distribution vanishes. Equation~\eqref{eq:fB} naturally includes the onset point of Quarkyonic Matter for which $\kbu=0$ and $E(\ksh)=\muB$. The suppressed baryon distribution in the deeper Fermi sea allows baryons to become relativistic without causing a rapid increase in baryon density. This represents a crucial mechanism for achieving a stiff equation of state. 
For neutron star physics, this transition to Quarkyonic Matter must take place at densities comparable to nuclear matter density. As noted in~\cite{Drischler:2020fvz}, the observed small tidal deformability from GW170817~\cite{LIGOScientific:2017vwq} and constraints from chiral effective field theory~\cite{Drischler:2020hwi} suggest that the onset of this rapid stiffening should occur at densities around $1.5$–$1.8$ times nuclear saturation density.

The shell structure of $\fB$ has further intriguing consequences. A first and direct observation is that contrary to the ideal Fermi gas, the energy of the state with the highest occupied momentum, here $\ksh$, is not identical to the baryon chemical potential anymore. The baryon chemical potential is determined as the change in energy density when the baryon density is increased, $\mu_B = \partial\varepsilon/\partial n_B$. In Quarkyonic Matter, if we add a baryon at $\ksh$, the energy of the system does not only increase by $E(\ksh)$. The increase in baryon density also pushes the edge of the bulk part to larger values, $\kbu \to \kbu + \delta\kbu$. As a consequence, the weight of baryon density in $\delta\kbu$ is pushed to higher momenta thus increasing further the energy density. The baryon chemical potential in the Quarkyonic Matter phase at a given baryon density is thus larger than the baryon chemical potential in the ideal Fermi gas with Fermi momentum $\ksh$. The pressure is obtained from $p = -\varepsilon + \mu_B n_B$, with the physical baryon chemical potential $\mu_B = \partial\varepsilon/\partial n_B$. Therefore, the pressure in Quarkyonic Matter is larger than in an ideal Fermi gas of nucleons for a given $\nB$. This modification of the baryon chemical potential is one of the important elements toward solving the ``hyperon puzzle'' as discussed in~\cite{Fujimoto:2024doc}.

It is the goal of this paper to extend the above description of Quarkyonic Matter to non-zero temperature. Here, the determination of the entropy is crucial and, as we will discuss in the next section, not straightforward.  Once the entropy density $s$ is properly derived for Quarkyonic Matter we can calculate the pressure via $p = -\varepsilon + \mu_B n_B + T s$, where again the physical baryon chemical potential is $\mu_B = (\partial\varepsilon/\partial n_B)\big|_s$. Now, in addition we find the physical temperature $T$ from $1/T = \beta = (\partial s/\partial\varepsilon)\big|_{n_B}$. It quantifies the change in entropy density when energy density is increased and the baryon density is held fixed. Similar to the baryon chemical potential at zero $T$, we expect that at non-zero $T$ the physical baryon chemical potential and the physical temperature are different from the corresponding Lagrange multipliers in the Fermi-Dirac distribution. In order to quantify this difference we first have to understand the entropy of Quarkyonic Matter. A naive application of the usual expression for entropy in the grand canonical ensemble suffers from an ill-defined zero temperature limit, as we will outline in section \ref{sec:entropy}.

For solving this issue, the saturation of the quark phase-space density plays an important role~\cite{Kojo:2021ugu, McLerran:2024rvk}. As a result, quarks are subject to the Pauli exclusion principle and follow the Fermi-Dirac statistics. At the same time, the quarks remain confined inside the nucleons and do not follow a thermal distribution themselves. They simply follow the nucleon thermodynamics. Let us assume an $N$-body system of nucleons, where each nucleon is a composite object of $\Nc$ quarks. Then the $N$-body nucleon wavefunction is properly anti-symmetrized via the Slater determinant. At high enough nucleon density, an additional constraint on the quarks becomes explicit by requiring anti-symmetry under the exchange of two quarks of the same color inside two nucleons. 
This additional constraint limits the available Fock space for the $N$-nucleon wavefunctions to those that are also antisymmetric in the quark states. At the same time, the one-particle nucleon states remain unchanged as the quark saturation, or quark Pauli exclusion criterion, acts as an external constraint.

As a consequence, the density of states of nucleons $g(k)$ is diluted as some states of the ideal gas are not available to nucleons in the Quarkyonic Matter regime due to the quark saturation constraint. As discussed, this has been observed in zero-$T$ Quarkyonic Matter~\cite{Fujimoto:2023mzy}, where for momenta below $\kbu$ we have $f_B(k\leq\kbu, T =0)=1/\Nc^3$. We can now understand this behavior as a consequence of the decreased density of states of nucleons below $\kbu$. In this sense, the nucleon states are not underoccupied. They are fully occupied  but there are simply less states available. We may then understand the baryon distribution function as the product of the density of states and the thermal distribution.

While the finite temperature extension of Quarkyonic Matter has been discussed in the literature~\cite{Sen:2020peq, Folias:2025zmu}, the analysis has largely remained ad hoc, in the sense that thermal effects were incorporated through phenomenological prescriptions rather than being systematically derived. In particular, the treatment of temperature has not been firmly rooted in a microscopic statistical-mechanical framework consistent with the underlying degrees of freedom of the system.
In this work, we go beyond such approaches by formulating the finite-temperature extension on a more fundamental footing. Starting from a well-defined statistical ensemble, we derive the thermodynamic quantities and distribution functions in a manner that is consistent with the underlying quark degrees of freedom and their constraints. This allows us to incorporate thermal effects in a controlled and systematic way, providing a more robust foundation for the study of Quarkyonic Matter at finite temperature.

In this work we will first introduce the problem of the entropy density in Quarkyonic Matter and the zero temperature limit together with its solution in section \ref{sec:entropy}. Next, we formally derive the grand canonical ensemble in the IdylliQ model framework, Section \ref{sec:grand}, and for a restricted Fock space, Section \ref{sec:derivation}. In Section \ref{sec:densofstate}, we derive the density of states in Quarkyonic Matter from maximizing the entropy density under inequality constraints. The behavior of the physical temperature and the physical baryon chemical potential is discussed in Section \ref{sec:6}.

We work in natural units $\hbar = c = k_B = 1$. For all numerical results shown in this article we consider $\Nc=3$ colors, $\Lambda=0.3$~GeV and $d=4$ degrees of freedom (assuming isospin symmetry) with mass $M_N=0.94$~GeV.

\section{The problem of non-vanishing entropy at zero temperature and its solution}
\label{sec:entropy}

In the dual picture of Quarkyonic Matter, the thermodynamic quantities are simultaneously described via the baryonic and the quark degrees of freedom. At zero temperature this has been worked out explicitly~\cite{Fujimoto:2023mzy}. For the baryon density, for $d$ degenerate baryon degrees of freedom, this translates to
\begin{equation}
  \label{eq:duality_n}
   \nB = d \int \!\! \frac{d^3 \bk}{(2\pi)^3}\, \fB(k)
   = d \int \!\! \frac{d^3 \bq}{(2\pi)^3}\, \fQ(q) \, .
\end{equation}
The energy density is similarly expressed via
\begin{align}
    \label{eq:duality_e}
    \eB [\fB] &= d \int \!\! \frac{d^3 \bk}{(2\pi)^3}\, E_{\mathrm{B}}(k) \,\fB(k)  \,,\\
    \eQ [\fQ] &= d \int \!\! \frac{d^3 \bq}{(2\pi)^3}\, E_{\mathrm{Q}}(q) \,N_c \,\fQ(q) \, . 
    \label{eq:eden_q}
\end{align}
In these expressions, the dispersion relation for a baryon is fixed as
\begin{equation}
    E_{\mathrm{B}}(k) = \sqrt{k^2 + M^2}\,,
\end{equation}
where $M$ is the mass of the baryon. From the duality relation $\varepsilon = \eB [\fB] = \eQ [\fQ]$, the dispersion relation for a quark is found to be
\begin{equation}
    E_{\mathrm{Q}}(q) = \sqrt{q^2 + M_{\mathrm{Q}}^2} \left[1 - \frac{2 \Lambda^2}{q^2 + M_{\mathrm{Q}}^2} - \frac{M_{\mathrm{Q}}^2 \Lambda^2}{(q^2 + M_{\mathrm{Q}}^2)^2}\right]\,,
    \label{eq:EQ}
\end{equation}
where $M_{\mathrm{Q}} = M/\Nc$ is the constituent quark mass. 

Going to non-zero temperature, we need to include the entropy as thermodynamic quantity, while in the zero-temperature limit the expression of the entropy has to vanish. 
For a degenerate Fermi gas with distribution function $f$, the standard textbook expression for the entropy density is given by, e.g.~\cite{Landau1980}, 
\begin{equation}
    s = -\,d \int \!\! \frac{d^3 \bk}{(2\pi)^3} \, \left\{ f(k) \ln f(k) + \left[1 - f(k) \right] \ln \left[1 - f(k) \right] \right\} \, ,
    \label{eq:snaiv}
\end{equation}
which vanishes in the zero temperature limit if $f$ is the Fermi-Dirac distribution function, because we have either $\fFD(T=0)=0$ or $\fFD(T=0)=1$. If, however, we evaluate the entropy density Eq.~(\ref{eq:snaiv}) with $f$ being the zero temperature baryon distribution function from Eq.~(\ref{eq:fB}), we see that the bulk part, which comes with a factor $1/\Nc^3$, contributes a non-zero  entropy. This is in strict contradiction to the third law of thermodynamics and must therefore be cured. 

As motivated in the introduction, the additional constraint from the Pauli exclusion principle applied to quarks limits the number of physically allowed many-body states in the system to a subspace of the Fock space associated with an ideal Fermi gas. There are fewer states to be occupied by baryons in the bulk part of Quarkyonic Matter, i.e.\ where the quark occupation constraint is active, compared to an ideal Fermi gas. This reduction of available states can be taken into account by the appropriate definition of the density of states $g(k)$. Then we may write the baryon distribution function as the product of the density of states and the thermal Fermi-Dirac distribution function associated with the ideal Fermi gas via
\begin{equation}
    \fB(k) = g(k) \fFD(k)\, .
    \label{eq:gfFD}
\end{equation}
The density of states $g(k)$ counts the available states in the phase space volume $d^3 k$ and is directly connected to the quark saturation constraint below $\qbu$, i.e.\ $\fQ(q\leq\qbu) \leq 1$. As a consequence, the density of states $g(k)$ of baryons in momentum space is suppressed because some states of the ideal Fermi gas are not allowed to baryons in Quarkyonic Matter. 

In this picture, one may interpret the above zero temperature solution for the baryon distribution function in Eq.~(\ref{eq:fB}) as follows: The Fermi-Dirac distribution at zero temperature is $\fFD(k) = \Theta(\ksh - k)$, and the density of states is $g(k) = \frac{1}{N_c^3}\Theta(\kbu - k) + \Theta(k - \kbu)$. This implies that the baryon states in the bulk part of the distribution, i.e. below $\kbu$, are not underoccupied but there are simply fewer states for the baryons available to be occupied. Nonetheless, all available states are occupied, which corresponds to a fully ordered system with zero entropy. 

Instead of the expression in Eq.~(\ref{eq:snaiv}) above, the entropy density of Quarkyonic Matter in the thermodynamic limit for $d$ degrees of freedom should be given by 
\begin{equation}
    \label{eq:sFD}
    s = - d \int \!\! \frac{d^3 \boldsymbol{k}}{(2\pi)^3} \, g(k)\left\{\fFD(k) \ln \fFD(k) + \left[1 - \fFD(k) \right] \ln \left[1 - \fFD(k) \right] \right\} \,.\\
\end{equation}
In this formulation the Fermi-Dirac distribution contains two Lagrange multipliers, $\hat T$ and $\hat \mu$, which control the energy and baryon number in the system, but as mentioned, are not the physical temperature and not the physical baryon chemical potential.

In the remainder of this paper, we put the suggested solution on solid grounds. We first outline how this entropy density expression is found from a straightforward modification of the ideal gas grand canonical ensemble, before we derive it rigorously from the Quarkyonic Matter grand canonical ensemble. We will see that this picture naturally emerges when one considers the saturation condition for the quark distribution function as an external condition in the statistical mechanics treatment. Finally, we determine the density of states from maximizing the entropy under inequality constraints and outline how the physical $T$ and $\muB$ can be deduced.

\section{Grand canonical ensemble in the IdylliQ picture}
\label{sec:grand}

In this section, we give an intuitive picture of how the grand canonical ensemble formulation of a quantum gas of non-interacting baryons can be modified in order to take into account the additional constraint of quark saturation. To simplify the discussion, we omit the degeneracy factor.

Consider a grand canonical ensemble enclosed in a volume $V$, where the energy and the baryon density are controlled by the parameters $\hat \beta = 1/\hat T$ and $\hat \mu$. In the ideal Fermi gas, these parameters would be the inverse temperature and the chemical potential.
The microscopic states can be specified by the set of parameters $(V, N, l)$, where $N$ is the particle number, and $l$ is the quantum number representing the quantum state.
As in the textbook treatment (e.g., see Ref.~\cite{Kubo:1965}), the probability of the system having the microscopic state $(V, N, l)$ is given by the grand canonical distribution
\begin{equation}
    \prob_{Nl}(V, \betaid, \muid) = \frac{1}{\Xi} e^{-\betaid(E_{VNl}  -\muid N)}\,,
\end{equation}
where $E_{VNl}$ is the energy of the quantum state $l$, and $\Xi$ is the grand partition function defined by
\begin{equation}
    \Xi(V, \betaid, \muid) \equiv \sum_{N, l} e^{-\betaid (E_{VNl} - \muid N)}\,.
\end{equation}
It is summed over all possible states $(V, N, l)$. In the following, we will suppress the $V$-dependence of $E_{VNl}$ because $V$ is fixed.

Let the one-particle states of a baryon be $\{ |\psi_i \rangle \}$ ($i = 1, 2, \ldots$).
We label the states so that the energy $E_i$ of each state $|\psi_i\rangle$ is in ascending order, i.e., $E_1 < E_2 < \cdots $.
Since baryons are non-interacting in the current setup, the quantum state $l$ of the total system is specified by a set of occupation numbers of one-particle states $\bn = (n_1, n_2, \ldots, n_i, \ldots)$, where $n_i$ is the occupation number of the one-particle state $|\psi_i\rangle$, and $n_i$ only takes the values $0$ or $1$ as a baryon is a fermion obeying the Pauli exclusion principle.
Then, the particle number and the energy of the multi-particle state $|\bn \rangle \equiv |n_1, n_2, \ldots \rangle$ are
\begin{equation}
    N = \sum_i  n_i\,,\qquad
    E_{\bn} = \sum_i E_i n_i\,.
    \label{eq:fixedNE}
\end{equation}

\subsection{Modification by the Pauli exclusion principle for quarks\label{sec:notation}}

If we consider now the Pauli exclusion principle for quarks in addition to that for baryons, we have to modify the grand canonical ensemble formulation. In particular, we need to eliminate some of the low-lying baryonic states $|\psi_i\rangle$ that are below a certain level $i < i_{\rm bu}$ and which cannot be occupied due to the Pauli exclusion principle on their constituent quarks. We characterize the ideal-gas states by the factor $\bg = (g_1, g_2, \ldots, g_i, \ldots)$. In the bulk, i.e.~below $i \leq i_{\rm bu}$, the allowed states are characterized by $g_i = 1$ and the forbidden states are characterized by $g_i = 0$. Above $\kbu$, i.e.~for $i > i_{\rm bu}$, all states are allowed, therefore $g_i = 1$. In this sense, the factor $g_i$ acts as a projector onto the allowed states. We note that even in the dual picture, the quantum states $|\psi_i \rangle$ of baryons remain the same since they are the eigenstates of the baryonic Hamiltonian, and the quark Pauli exclusion principle is an external condition that is not intrinsic to the baryonic description. Therefore, in the derivation below, the index $i$ runs over all the quantum states of baryons as given in the ideal gas picture. This $g_i$ factor will lead to the non-trivial density of states $g(k)$ in the thermodynamic limit as will be explained below.

In the following, we consider the case in which the microscopic label $l$ is given by the element-wise product of $\bg$ and $\bn$, i.e.~$\bg \odot \bn \equiv (g_1 n_1, g_2 n_2, \ldots)$. This means that we consider the multi-particle state $|\bg \odot \bn \rangle \equiv |g_1 n_1, g_2 n_2, \ldots \rangle$. The factor $g_i$ takes only the values $0$ or $1$, so it satisfies the property $g_i^2 = g_i$. The particle number and the energy of the multi-particle state $|\bg \odot \bn \rangle$ are
\begin{equation}
    N = \sum_i g_i n_i\,,\qquad
    E_{\bg \odot \bn} = \sum_i g_i E_i n_i\,.
    \label{eq:NgnEgn}
\end{equation}

In this setup, once the $\bg$ and $\bn$ are specified, $N$ is given automatically. 
Therefore, the sum $\sum_{N,l}$ becomes the sum over $\bg \odot \bn$:
\begin{equation}
    \sum_{N, l} = \sum_{\bg \odot \bn} \equiv \prod_i \sum_{n_i = 0}^{g_i}\,.
\end{equation}
The grand canonical partition function is
\begin{equation}
    \Xi = \sum_{\bg \odot \bn} e^{-\betaid(E_{\bg \odot \bn} - \muid N)} = \prod_{i} \left[1 + g_i e^{-\betaid (E_i - \muid )}\right]\,.
    \label{eq:Xi}
\end{equation}
The probability of a microscopic state $| \bg \odot \bn \rangle$ to be realized is
\begin{equation}
    \prob_{\bg \odot\bn} = \frac{1}{\Xi} e^{-\betaid(E_{\bg \odot \bn}  -\muid N)}\,.
\end{equation}
The mean number of particles in the state $i$ is
\begin{align}
    \langle n_i \rangle 
    &= \sum_{\bg \odot \bn} n_i \prob_{\bg \odot \bn} \,, \notag \\
    &= \frac{1}{\Xi}\sum_{n_i = 0}^{g_i} n_i e^{-\betaid (E_i - \muid )g_i n_i} \prod_{j \neq i} \sum_{n_j = 0}^{g_j} e^{-\betaid (E_j - \muid )g_j n_j}\,, \notag\\
    &=
    \frac{g_i}{e^{\betaid(E_i - \muid)} + 1}\,.
    \label{eq:nmean}
\end{align}
Using the average occupation number $\langle n_i \rangle$, and because $g_i\in\{0,1\}$, the partition function can be rewritten in the form
\begin{equation}
    \Xi = \prod_{i} \frac{1}{1 - \langle n_i \rangle}\,.
    \label{eq:xini}
\end{equation}
The number density $n$ and energy density $\varepsilon$ are
\begin{align}
     n
    &= \frac{\langle N \rangle}{V}
     = \frac{1}{V} \sum_{\bg \odot \bn} N \prob_{\bg \odot \bn} 
    = \frac{1}{V} \sum_{\bg \odot \bn} \left(\sum_i g_i n_i \right) \prob_{\bg \odot \bn} 
     = \frac{1}{V} \sum_i g_i \langle n_i \rangle\,, 
\label{eq:nsum} \\
    \varepsilon
    &= \frac{\langle E_{\bg \odot \bn} \rangle}{V}
    = \frac{1}{V} \sum_{\bg \odot \bn} E_{\bg \odot \bn} \prob_{\bg \odot \bn} 
     = \frac{1}{V} \sum_{\bg \odot \bn} \left(\sum_i E_i g_i n_i \right) \prob_{\bg \odot \bn} 
     = \frac{1}{V} \sum_i g_i E_i \langle n_i \rangle\,.
    \label{eq:esum}
\end{align}
The entropy is given by the expression of the Gibbs entropy $\langle S \rangle = - \sum_{Nl} \prob_{Nl} \ln \prob_{Nl}$.
Accordingly, the entropy density is
\begin{align}
    s = \frac{\langle S \rangle}{V}
    &= - \frac{1}{V} \sum_{\bg \odot \bn} \prob_{\bg \odot \bn} \ln \prob_{\bg \odot \bn} \,, \notag \\
    &=  \frac{1}{V} \ln \Xi + \frac{1}{V} \sum_i \betaid g_i (E_i - \muid ) \langle n_i \rangle \,,\notag\\
    &= -\frac{1}{V} \sum_i g_i \left[\langle n_i \rangle \ln \langle n_i \rangle + \left(1 - \langle n_i \rangle\right) \ln \left(1 - \langle n_i \rangle\right) \right]\,.
    \label{eq:ssum}
\end{align}
In the derivation we have used the relation $\betaid (E_i - \muid) = \ln [\langle n_i \rangle / (1 - \langle n_i \rangle)]$ and Eq.~\eqref{eq:xini}.

\subsection{Thermodynamic limit}

Now that we have expressed the thermodynamic quantities, we can take the thermodynamic limit, i.e.~$V \to \infty$, $N \to \infty$, and $N/V = \mathrm{fixed}$.
The index $i$ that labels the one-particle state $|\psi_i\rangle$ corresponds to the momentum $\boldsymbol{k}$.
Thus, in the thermodynamic limit, assuming isotropy in momentum-space, the sum over $i$ becomes
\begin{equation}
    \sum_i \to  \frac{1}{\Delta k} \int d^3 \boldsymbol{k} = \frac{V}{2\pi^2} \int k^2 dk\,, \qquad
    \Delta k = \frac{(2\pi)^3 }{V}\,.
    \label{eq:levelspac}
\end{equation}
The aforementioned $g_i$-factor will turn into the non-trivial single-particle density of states $g(k)$ in  momentum space defined by
\begin{equation}
    \frac{k^2}{2\pi^2} g(k) \delta k \equiv \frac{1}{V} \sum_{k < k_i < k +\delta k} g_i\,,
    \label{eq:dos}
\end{equation}
where $k_i$ is the momentum of a particle of energy $E_i$, i.e.~$k_i = \sqrt{E_i^2 - M^2}$, and $\delta k$ is an infinitesimal momentum variation. Since $g_i \in \{0,1\}$ acts as a projector onto the allowed single-particle states, any function $f(g_i)$ appearing under the sum obeys the identity $g_i f(g_i) = g_i f(1)$. Consequently, in the thermodynamic limit, expressions of the form $\sum_i g_i f(g_i)$ reduce to integrals weighted by the effective density of states $g(k)$
\begin{equation}
    \frac{k^2}{2\pi^2} g(k) f(1)\delta k = \frac{1}{V} \sum_{k < k_i < k +\delta k} g_i f(g_i)\,.
    \label{eq:densandfunc}
\end{equation}

In the thermodynamic limit, the thermodynamic quantities~(\ref{eq:nsum})--(\ref{eq:ssum}) become
\begin{align}
    \nB &= \int \!\! \frac{d^3 \boldsymbol{k}}{(2\pi)^3} \, g(k) \fFD(k) \equiv \int \!\! \frac{d^3 \boldsymbol{k}}{(2\pi)^3} \,\fB(k)\,,
    \label{eq:n}\\
    \varepsilon &= \int \!\! \frac{d^3 \boldsymbol{k}}{(2\pi)^3} \, E_{\mathrm{B}}(k) g(k) \fFD(k) \equiv \int \!\! \frac{d^3 \boldsymbol{k}}{(2\pi)^3} \, E_{\mathrm{B}}(k) \fB(k)\,,
    \label{eq:e}\\
    s &= -\int \!\! \frac{d^3 \boldsymbol{k}}{(2\pi)^3} \, g(k)\left\{ \fFD(k) \ln \fFD(k) + \left[1 - \fFD(k) \right] \ln \left[1 - \fFD(k) \right] \right\} \,,
    \label{eq:s}
\end{align}
where the Fermi-Dirac distribution function $\fFD(k)$, corresponding to Eq.~\eqref{eq:nmean}, is
\begin{equation}
    \fFD(k) \equiv \frac{1}{e^{\betaid[E_{\mathrm{B}}(k) - \muid]}+ 1}\,.
\end{equation}
We find that the baryon distribution function is given by the product 
\begin{equation}
    \fB(k) = g(k)\,\fFD(k)\,,
\end{equation}
as anticipated in Eq.~\eqref{eq:gfFD}.

Since both the baryon density $n_B$ and the energy density $\varepsilon$ depend linearly on the distribution function, the zero-temperature expressions of the IdylliQ model could be written solely in terms of $\fB(k)$ without explicitly distinguishing between the density of states $g(k)$ and the occupation factor $\fFD(k)$. The entropy density, however, depends nonlinearly on $\fFD(k)$ and is explicitly weighted by the density of states $g(k)$. Therefore, the separation between density of states and thermal occupation becomes essential at non-zero temperature. This observation clarifies the difficulty encountered in a naive extension of the zero-temperature distribution to non-zero temperature.

\subsection{The zero temperature limit}

Let us verify that the baryon density, energy density, and entropy density obtained from the modified grand canonical ensemble reproduce the zero-temperature results derived previously in
Refs.~\cite{Fujimoto:2023mzy, Fujimoto:2024doc}. In the zero-temperature limit, the Fermi--Dirac distribution reduces to a
step function, $\fFD(k) \to \Theta[\muid - E_{\mathrm{B}}(k)]$.
Substituting this expression into Eq.~\eqref{eq:s}, one immediately finds that the entropy density vanishes, $s=0$, as required by the third law of thermodynamics.

Moreover, inserting the step function into
Eqs.~\eqref{eq:n} and~\eqref{eq:e} reproduces the
previous zero-temperature expressions,
Eqs.~\eqref{eq:duality_n} and~\eqref{eq:duality_e}
(for $d=1$).

\begin{align}
    \nB &= \int \!\! \frac{d^3 \boldsymbol{k}}{(2\pi)^3} \, g(k) \Theta[\muid - E_{\mathrm{B}}(k)]\,,\\
    \varepsilon &= \int \!\! \frac{d^3 \boldsymbol{k}}{(2\pi)^3} \, g(k) E_{\mathrm{B}}(k) \Theta[\muid - E_{\mathrm{B}}(k)]\,,
\end{align}
where $\muid$ is related to $\ksh$ in Eq.~\eqref{eq:fB} via $\muid = \sqrt{\ksh^2 + M^2}$, but does not coincide with the physical baryon chemical potential. The density of states $g(k)$ in the IdylliQ model can be identified as
\begin{equation}
    \label{eq:g}
    g(k) = \frac{1}{N_c^3} \Theta(\kbu - k) + \Theta(k - \kbu)\,,
\end{equation}
where $\kbu$ is related to the bulk energy scale $E_{i_{\rm bu}}$. We will discuss its derivation, as well as its non-zero-temperature counterpart, below.

With this form of $g(k)$, the baryon distribution function is indeed given by the product $f_B(k)=g(k)\fFD(k)$, as stated in Eq.~\eqref{eq:gfFD}. All thermodynamic quantities therefore reproduce the correct zero-temperature limit.

\subsection{Physical temperature}

As we have discussed in the introduction, in Quarkyonic Matter the parameters in the distribution function controlling the mean particle number and mean energy, $\hat\mu$ and $\hat\beta$, do not correspond to the physical baryon chemical potential $\muB$ and the inverse physical temperature $1/T$. 
The difference between $\hat\mu$ and $\muB$ has been elaborated for the zero temperature case in the introduction, and its physical understanding does not change at non-zero temperature. 

The physical temperature corresponds to the effective temperature emerging from the system itself as it is exposed to the additional constraint of quark saturation. It can be determined from $1/T = (\partial s/\partial \varepsilon)\big|_{\nB}$. We can already get a physical intuition of what to expect for this physical temperature. We discussed that the integral duality relation in Eq.~(\ref{eq:duality}) can be inverted via the differential operator $\hat{L}_q$ in Eq.~(\ref{eq:diffOpL}). This leads to the interesting consequence that the low-momentum part of the distribution function becomes medium independent (this low-momentum part is limited by $\kbu$, which itself depends on the medium). As a consequence, thermal excitations of baryons from deep inside the Fermi sea are blocked, and thus the effective, physical temperature inside the medium is reduced. 

Let us  consider a thermal excitation around the (smeared out) Fermi surface: if a baryon is excited to a higher momentum, the weight of the quark distribution at lower momentum is reduced and as a result $\kbu$ will move down to $\kbu - \delta\kbu$. As a consequence, first, the change in energy is not as large as the energy associated to the simple transfer of one baryon to higher momentum, and second, the increase in entropy is enormous for the momentum region between $\kbu$ and $\kbu-\delta\kbu$ as it changes from being suppressed by $1/\Nc^3$ to not being suppressed. These three aspects, the locking of the non-thermal low-momentum region forbidding thermal excitations from the deep Fermi sea, the reduced change in energy for the thermal excitation of baryons around the Fermi surface and the enormous entropy increase associated with this thermal excitation, lead us to expect that in Quarkyonic Matter the physical temperature is significantly smaller than $\hat T$. We will verify this expectation once we will have derived the density of states at non-zero temperature in section \ref{sec:densityofstatesT}.

\subsection{Physical pressure, the grand potential and the partition function\label{sec:TmuOmega}}

After having obtained the expressions for energy density, baryon density and entropy density in the Quarkyonic Matter phase at non-zero temperature, we will now discuss another important quantity, the pressure. The pressure quantifies the response of the system to changes in the volume at constant physical temperature $T$ and constant physical baryon chemical potential $\muB$. In the standard formulation of thermodynamics, all momentum states see the same total volume $V$. From Eq.(\ref{eq:dos}) it is clear that the Quarkyonic Matter the effective volume of the system $V_{\rm eff} = Vg(k)$ is momentum dependent. Therefore, changing the physical volume does not change all states uniformly. Then, a naive identification with the logarithm of the partition function in Eq.~(\ref{eq:Xi}) (in the thermodynamic limit) is not possible:
\begin{equation}
   \hat{p} = \frac{\hat{T}}{V}\ln \Xi = \hat{T} s + \muid \nB - \varepsilon \neq p \,. 
\end{equation}
Instead, the physical pressure follows from the Euler relation 
\begin{equation}
  p = Ts + \muB \nB - \varepsilon
  \label{eq:PressEuler}
\end{equation}
which is fundamentally derived from the first law of thermodynamics. We can relate the physical pressure $p$ to $\hat p$ via
\begin{equation}
    \frac{p}{T}-\frac{\hat p}{\hat T} = \varepsilon \left( \frac{1}{\hat T} - \frac{1}{T}\right) - \nB\left(\frac{\hat \mu}{\hat T} - \frac{\muB}{T}\right) \,.
\end{equation}
For practical applications, the physical pressure should best be calculated from the Euler relation in Eq.~(\ref{eq:PressEuler}).

\section{Derivation of the grand canonical ensemble for a restricted Fock space\label{sec:derivation}}

In this section, we derive the grand canonical ensemble from the micro canonical ensemble incorporating the $g_i$ factor introduced above to describe the suppression of non-physical states due to the quark Pauli exclusion principle. With the help of $g_i$, we project our system onto the allowed states. Our main goal is to derive the expression for the entropy density in Eq.~(\ref{eq:s}) on this restricted Fock space. The splitting of $\fB$ into the product $\fB(k) = g(k)\fFD(k)$ emerges naturally. 

We start with Boltzmann's formula for the entropy
\begin{equation}
    \label{eq:Boltzmann}
    S = \ln \mathcal{W} (E, N)\,,
\end{equation}
where, for a system composed of $N$ particles, $\mathcal{W} (E, N)$ is the total number of possible quantum states in the system having an energy eigenvalue that lies between $E$ and $E+\delta E$, where the value of the small energy variation $\delta E$ is not important. 
In applying Boltzmann's formula, the counting in $\mathcal{W}$ is crucial. 
To this end, we define the statistical weight $W(\bn) = W(n_1, n_2, \ldots, n_i, \ldots)$ as the number of quantum states of the system corresponding to the set of occupation numbers $\bn = (n_1, n_2, \ldots, n_i, \ldots)$.
The Fermi-Dirac statistics is characterized by the following statistical weight:
\begin{equation}
    \label{eq:stat}
    W(\bn) =
    \begin{cases}
        1 & \text{(all $n_i$ are $n_i = 0$ or $1$)} \\
        0 & \text{($n_i > 1$ for any of the $n_i$)}\,.
    \end{cases}
\end{equation}
Then, $\mathcal{W} (E, N)$ is given by the sum of $W(\bn)$ as
\begin{equation}
    \label{eq:multi}
    \mathcal{W}(E, N) = {\sum_{\bn}}' W(\bn)\,,
\end{equation}
where the sum ${\sum_{\bn}}'$ is taken over all possible sets $\bn$ subject to the constraints of fixed $N$ and $E$ as in Eq.~\eqref{eq:fixedNE}.

There are two known methods for evaluating $\mathcal{W}(E,N)$ in Eq.~\eqref{eq:multi} depending on how one defines the thermodynamic equilibrium: either via the most probable distribution or via the mean distribution. For a system with a large number of particles $N$ the two definitions will coincide by virtue of the central limit theorem. It is more common to evaluate the equilibrium distribution from the maximum entropy principle, which corresponds to evaluating the most probable distribution. This method is briefly reviewed in Appendix~\ref{sec:dirac}. It relies on the grouping of states into cells. Since the $g_i$ factor acts on each sub-level of the grouped states, there remains an ambiguity in how this factor $g_i$ should be handled in the grouping of the states into cells.  

In what follows we will therefore directly evaluate $\mathcal{W}$ from the microscopic states by using the generating function method and the saddle point method. 
This is known as the method of the mean values, or as Darwin-Fowler method~\cite{Darwin:1922, Fowler:1926}, which was first devised as a method to rigorously derive the canonical ensemble without using Stirling's approximation. We follow its presentation from standard textbooks~\cite{Schrodinger:1952, terHaar:1995} and adapt it to the restricted Fock space.

We will calculate the mean number of particles in the state $i$, $\langle n_i \rangle$, and the total multiplicity of the system, $\mathcal{W}$, under the condition that the total number of particles $N$ and the energy $E$ are fixed, i.~e.
\begin{equation}
    \label{eq:fixedNE_g}
    N = \sum_i g_i n_i = \sum_i g_i \langle n_i \rangle \,,\quad
    E = \sum_i g_i E_i n_i = \sum_i g_i E_i \langle n_i \rangle \,.
\end{equation}
In terms of the statistical weight Eq.~\eqref{eq:stat}, we can write $\langle n_i \rangle$ and $\mathcal{W}$ as (see section~\ref{sec:notation} for notation) 
\begin{align}
    \langle n_i \rangle &= \frac{1}{\mathcal{W}} {\sum_{\bg \odot \bn}}' (g_i n_i) W(\bg \odot \bn)\,,\label{eq:meanni}\\
    \mathcal{W}(N, E) &= {\sum_{\bg \odot \bn}}' W(\bg \odot \bn)\,,
\end{align}
where the sum ${\sum}'$ is taken over all possible sets of $\bg \odot \bn$ under the conditions in Eq.~\eqref{eq:fixedNE_g}. Now, the statistical weight can be written in terms of the product of each state
\begin{equation}
    W(\bg \odot \bn) = \prod_i \eta (g_i n_i)\,,
\end{equation}
where in the Fermi-Dirac statistics one has $\eta(n_i)$ defined as
\begin{equation}
    \eta(n_i)
    =
    \begin{cases}
        1 & (n_i = 0 \text{ or } 1)\,,\\ 
        0 & (n_i > 1)\,.
    \end{cases}
\end{equation}

The generating function is given as
\begin{align}
    G(z, w; \bE, \bg) \equiv \sum_{\bg \odot \bn} W(\bg \odot \bn) \, z^{\sum_i g_i n_i} \, w^{\sum_i g_i E_i n_i}\,,
    \label{eq:charF}
\end{align}
where we define the energy corresponding to $\bn$ as $\bE \equiv (E_1, E_2, \ldots)$.
In this formula, the sum is taken over all possible values of $\bg \odot \bn$ without any restrictions, and we do not impose the conditions in Eq.~\eqref{eq:fixedNE_g}.
By using the statistical factor $\eta$, one can rewrite the generating function as a product of functions for each quantum state :
\begin{align}
    \label{eq:geneG}
    G(z, w; \bE, \bg) &= \prod_i h(z w^{E_i}, g_i)\,,\\
    h(x, g_i) &\equiv \sum_{n_i = 0}^{g_i} \eta(g_i n_i) x^{g_i n_i} = 1 + g_i x^{g_i}=
1+g_i x, \qquad g_i\in\{0,1\}\,.
    \label{eq:geneh}
\end{align}
We now determine the statistical weight $\mathcal{W}(E,N)$ and the mean occupation
numbers $\langle n_i\rangle$ from the generating function
$G(z,w;\mathbf{E},\mathbf{g})$. To this end, the constraints of fixed total particle
number and energy, Eq.~\eqref{eq:fixedNE_g}, must be reimposed. This is achieved by
projecting onto the coefficients of $z^N w^E$ via complex contour integration. Specifically, we obtain the statistical weight as 
\begin{equation}
    \mathcal{W} = \oint \frac{dz}{2\pi i} \oint \frac{dw}{2\pi i}\, z^{-N-1} w^{-E-1} G(z, w; \bE, \bg)\,. \label{eq:W}
\end{equation}
The integration contours are taken as closed counterclockwise paths around the origin. The contour integrals act as coefficient extractors and enforce the constraints on $N$ and $E$. Likewise, the mean occupation number $\langle n_i\rangle$ is obtained by inserting the appropriate derivative with respect to $E_i$ inside the same projection,
\begin{equation}
    \langle n_i \rangle = \frac{1}{\mathcal{W}} \oint \frac{dz}{2\pi i} \oint \frac{dw}{2\pi i}\, z^{-N-1} w^{-E-1} \left(\frac{1}{\ln w} \frac{\partial}{\partial E_i}\right) G(z, w; \bE, \bg)\,.
    \label{eq:n_i}
\end{equation}
The quantities $z$ and $w$ should be understood as formal generating variables.
Expressions such as $w^{-E-1}$ do not carry physical dimensions; dimensional
consistency is restored after evaluating the contour integrals in the saddle-point
approximation, where $z$ and $w$ are identified with thermodynamic parameters. We proceed to evaluate these complex integrals by the saddle point method. Then $\mathcal{W}$ is approximated by
\begin{equation}
    \mathcal{W} \simeq z_0^{-N-1} w_0^{-E-1} G(z_0, w_0; \bE, \bg)\,.
    \label{eq:calW}
\end{equation}
Similarly, one can use the saddle point method for $\langle n_i \rangle$ in Eq.~\eqref{eq:n_i} and, by substituting Eqs.~\eqref{eq:geneG} and \eqref{eq:geneh} into the expression, one obtains
\begin{align}
    \langle n_i \rangle
    &\simeq \frac{1}{\mathcal{W}} z_0^{-N-1} w_0^{-E-1} \left(\frac{1}{\ln w_0} \frac{\partial}{\partial E_i}\right) G(z_0, w_0; \bE, \bg)\,, \notag \\
    &= \frac{1}{\ln w_0} \frac{1}{h(z_0 w_0^{E_i}, g_i)} \frac{\partial h(z_0 w_0^{E_i}, g_i)}{\partial E_i}\,,
    \label{eq:nbar}
\end{align}
where $z_0$ and $w_0$ are the saddle points. They are determined as the extrema of the integrand in Eq.~\eqref{eq:W} and given by the following equations :
\begin{equation}
    z \frac{\partial G}{\partial z} - (N+1) G = 0\,,\quad
    w \frac{\partial G}{\partial w} - (E+1) G = 0\,.
    \label{eq:z0w0}
\end{equation}
Since $N$ and $E$ are large, the approximations $N+1\approx N$ and in $E+1\approx E$ lead to the following expressions
\begin{align}
    N &= \left. \frac{\partial \ln G}{\partial \ln z}\right|_{z=z_0, w=w_0} = \sum_i \left. \frac{\partial \ln h}{\partial \ln z} \right|_{z=z_0, w=w_0}\,,\\
    E &= \left. \frac{\partial \ln G}{\partial \ln w} \right|_{z=z_0, w=w_0}
    = \sum_i \left. \frac{\partial \ln h}{\partial \ln w } \right|_{z=z_0, w=w_0}\,.
\end{align}
The mean number of particles $\langle n_i \rangle$ can be obtained from Eq.~\eqref{eq:nbar} as 
\begin{align}
    \langle n_i \rangle = \frac{g_i}{z_0^{-1} w_0^{- E_i} +1} \,. 
    \label{eq:ni}
\end{align}
Accordingly, the above equations can be written in terms of $\langle n_i \rangle$ as 
\begin{align}
    \label{equ:NandEStatMech}
    N = \sum_i g_i \langle n_i \rangle\,,\quad
    E = \sum_i g_i E_i \langle n_i \rangle\,.
\end{align}
The entropy can be obtained from Eq.~(\ref{eq:Boltzmann}) with $\mathcal{W}$ in Eq.~\eqref{eq:calW} as 
\begin{align}
    S = \ln \mathcal{W}
    & \simeq - (N+1) \ln z_0 - (E+1) \ln w_0 + \ln G(z_0, w_0; \bE, \bg)\,,\notag \\
    & \simeq - N \ln z_0 - E \ln w_0 + \sum_i \ln \left(1 + g_i z_0 w_0^{ E_i}\right) \,, \notag \\
    & = -\sum_i \left[g_i \langle n_i \rangle \ln \left(z_0 w_0^{E_i} \right) - g_i \ln \left(1 + z_0 w_0^{E_i}\right)\right]\,, \notag \\
    \label{equ:sStatMech}
    & = -\sum_i g_i \left[\langle n_i \rangle \ln \langle n_i \rangle + \left(1 - \langle n_i \rangle \right) \ln \left(1 - \langle n_i \rangle \right)\right]\,.
\end{align}
These thermodynamic quantities coincide with those obtained in the previous section, Eqs.~(\ref{eq:nsum})--(\ref{eq:ssum}). The expressions in the thermodynamic limit therefore reproduce Eqs.~(\ref{eq:n})--(\ref{eq:s}) including the particular form of the entropy density, which has the correct zero-temperature limit.

The undetermined parameters $w_0$ and $z_0$ in $\langle n_i \rangle$ are the saddle point values of the generating functional, which enforce the total particle number $N$ and energy $E$. They can be identified from comparison of Eq.~(\ref{eq:ni}) and Eq.~(\ref{eq:nmean}) with $w_0 = \exp{(-\betaid)}$ and $z_0 = \exp{(\betaid\muid)}$, but $1/\betaid$ and $\muid$ do not have to coincide with the emerging physical $T$ and $\muB$ of the Quarkyonic Matter thermodynamics.

We note in passing that the analysis we performed above can be considered as a further generalization of the Haldane fractional exclusion statistics (FES)~\cite{Haldane:1991xg, Wu:1994it}.
Here, we consider a type of statistics that goes in the opposite direction to FES, in the sense that the available single-particle states are further restricted.

\section{Determination of the density of states}
\label{sec:densofstate}

After having discussed the statistical mechanics of Quarkyonic Matter, we will now determine the density of states $g(k)$ by solving a constrained optimization problem.

\subsection{Method of Lagrange multipliers with inequality constraints}

We briefly review the method of Lagrange multipliers with inequality constraints, formulated through the Karush–Kuhn–Tucker (KKT) conditions \cite{boyd2004convex}. Consider the following constrained optimization problem: for a given function $f$ determine $\boldsymbol{x}$  that minimizes $f(\boldsymbol{x})$ is minimal subject to  :
\begin{align}
    g_i(\boldsymbol{x}) &= 0\,, \qquad (i = 1, \ldots, N)\,,\label{eq:EQC1}\\
    h_j(\boldsymbol{x}) &\leq 0\,, \qquad (j = 1, \ldots, M)\,.\label{eq:IEQC1}
\end{align}
The equality constraints in Eq. (\ref{eq:EQC1} are enforced by introducing Lagrange multipliers $\boldsymbol{\lambda} = (\lambda_1, \ldots, \lambda_N)$. The inequality constraints in Eq. (\ref{eq:IEQC1} are formally incorporated via non-negative multipliers $\boldsymbol{\mu} = (\mu_1, \ldots, \mu_M)$ leading to the following Lagrangian
\begin{equation}
    \mathcal{L}(\boldsymbol{x}, \boldsymbol{\lambda}, \boldsymbol{\mu}) = f(\boldsymbol{x})
    + \sum_{i=1}^{N} \lambda_i g_i(\boldsymbol{x})
    + \sum_{j=1}^{M} \mu_j h_j(\boldsymbol{x})\,.
\end{equation}
The sign of the terms proportional to $\mu_j$ depends on the direction of the inequality and on whether the problem is formulated as a minimization or maximization.

Let $\boldsymbol{x}^*$ denote a point that minimizes the function $f$ subject to the above constraints.
If the gradients 
$\nabla_{\boldsymbol{x}} g_i(\boldsymbol{x}^*)$ and
$\nabla_{\boldsymbol{x}} h_j(\boldsymbol{x}^*)$ are linearly independent, there exist multipliers $\boldsymbol{\lambda}^*$ and $\boldsymbol{\mu}^*$ such that
\begin{align}
    \nabla_x \mathcal{L}(\boldsymbol{x}^*, \boldsymbol{\lambda}^*, \boldsymbol{\mu}^*) &= 0 \,,
    \label{eq:KTT1}\\
    \mu^*_j h_j(\boldsymbol{x}^*) &= 0 \,,
    \label{eq:KTT2}\\
    \boldsymbol{\mu}^* &\geq 0\,.
    \label{eq:KTT3}
\end{align}
These relations constitute the KKT conditions~\cite{boyd2004convex}. 
In this general form, Eqs.~\eqref{eq:KTT1}–\eqref{eq:KTT3} are necessary conditions
for optimality. At the solution, some inequality constraints may be binding,
i.e.\ $h_j(\boldsymbol{x}^*)=0$, while others are non-binding,
$h_j(\boldsymbol{x}^*)<0$. For non-binding constraints the corresponding
multiplier vanishes, $\mu_j^*=0$, whereas binding constraints contribute with
$\mu_j^*\ge 0$. They can be treated analogously to equality constraints.

\subsection{Determination of $g(k)$ at zero temperature}

As a first application of the KKT optimization method, we consider the IdylliQ model at zero temperature and show that the density of states $g(k)$ is recovered as in Eq.~(\ref{eq:g}). We minimize the energy density $\varepsilon[g(k)]$ as a functional of $g(k)$ at a fixed baryon density. This variational problem with respect to $g(k)$ is subject to the following constraints :
\begin{align}
    \nB[g(k)] = n^*\,, \label{eq:cond_n}\\
    0 \leq \fFD(k) \leq 1\,, \label{eq:cond_fFD}\\
    0 \leq g(k) \leq 1\,, \label{eq:cond_g}\\
    0 \leq \fQ(q) \leq 1 \label{eq:cond_fQ}\,.
\end{align}
The conditions in Eqs.~\eqref{eq:cond_fFD} and~\eqref{eq:cond_g} together ensure that $0 \leq \fB(k) \leq 1$. The corresponding Lagrangian reads
\begin{align}
    \mathcal{L}[g(k)]
    &= \varepsilon[g(k)]
    - \lambda_n (\nB[g(k)] - n^*) \notag \\
    &\quad - \int_{\boldsymbol{k}} \lambda_{\fFD \ge 0}(k) \fFD(k)
    + \int_{\boldsymbol{k}} \lambda_{\fFD \leq 1}(k) \left[ \fFD(k) - 1 \right]\notag \\
    &\quad - \int_{\boldsymbol{k}} \lambda_{g \ge 0}(k) g(k)
    + \int_{\boldsymbol{k}} \lambda_{g \leq 1}(k) \left[ g(k) - 1 \right]\notag \\
    &\quad - \int_{\boldsymbol{q}} \lambda_{\fQ \geq 0}(q) \fQ(q)
    + \int_{\boldsymbol{q}} \lambda_{\fQ \leq 1}(q) \left[\fQ(q) - 1\right]\,.
    \label{eq:lagrangian}
\end{align}
From the statistical mechanical analysis at zero temperature, we use  $\fFD(k) = \Theta[\muid - E_{\mathrm{B}}(k)] = \Theta(\ksh - k)$ with $\ksh = \sqrt{\muid^2 - M^2}$.
This corresponds to the stationary point of $\mathcal{L}$ with respect to the variation of $\fFD$ in absence of the quark constraint Eq.~\eqref{eq:cond_fQ}. It fixes $\lambda_n = \muid$.

The first KKT condition, Eq.~(\ref{eq:KTT1}), yields the functional stationarity of the Lagrangian with respect to $g(k)$ as 
\begin{align}
    \nonumber
    0 = & \,\frac{\delta \mathcal{L}}{\delta g(k)} \\
        \label{eq:L_var}
    = & \left[E_{\mathrm{B}}(k) - \muid
    - \int_{\boldsymbol{q}}  \lambda_{\fQ \geq 0}(q) \,\varphi\left(\boldsymbol{q} - \frac{\boldsymbol{k}}{N_c}\right)
    + \int_{\boldsymbol{q}}  \lambda_{\fQ \leq 1}(q) \,\varphi\left(\boldsymbol{q} - \frac{\boldsymbol{k}}{N_c}\right) \right] \fFD(k) \\ \nonumber
    & \, - \lambda_{g \geq 0}(k)
    + \lambda_{g \leq 1}(k)\,.
\end{align}
The remaining KKT conditions Eq.~(\ref{eq:KTT2}) and Eq.~(\ref{eq:KTT3}) associated with the bounds on $g(k)$ become
\begin{align}
    \lambda_{g \geq 0}(k) g(k) &= 0 \,,
    \label{eq:KTT2g0}\\
    \lambda_{g \geq 0}(k) &\geq 0\,,
    \label{eq:KTT3g0}\\
    \lambda_{g \leq 1}(k) (g(k)-1) &= 0 \,,
    \label{eq:KTT2g1}\\
    \lambda_{g \leq 1}(k) &\geq 0\,.
    \label{eq:KTT3g1}
\end{align}
And similarly the constraint on the quark occupation reads
\begin{align}
    \lambda_{\fQ \geq 0}(q) \fQ(q) &= 0\,,
    \label{eq:KTT2fQ0}\\
    \lambda_{\fQ \geq 0}(q) &\geq 0\,,
    \label{eq:KTT3fQ0}\\
    \lambda_{\fQ \leq 1}(q) (\fQ(q) -1) &= 0\,,
    \label{eq:KTT2fQ1}\\
    \lambda_{\fQ \leq 1}(q) &\geq 0\,.
    \label{eq:KTT3fQ1}
\end{align}
Binding constraints correspond to strictly positive multipliers and saturated inequalities. The possible binding cases are therefore
\begin{enumerate}
    \item \label{cond:1} $\lambda_{g \geq 0}(k) > 0\,, \ g(k) = 0$\,,
    \item \label{cond:2} $\lambda_{g \leq 1}(k) > 0\,, \ g(k) = 1$\,,
    \item \label{cond:3} $\lambda_{\fQ \geq 0}(q) > 0\,, \ \fQ(q) = 0$\,,
    \item \label{cond:4} $\lambda_{\fQ \leq 1}(q) > 0\,, \ \fQ(q) = 1$\,.
\end{enumerate}
We assume that these binding conditions hold over finite, connected intervals in momentum space, with the corresponding quark momenta related by $q=k/N_c$. Outside these regions the constraints are non-binding and the associated multipliers vanish.

We now examine which of the above listed binding cases can be realized. Some of them are mutually exclusive. In particular, cases~\ref{cond:1} and~\ref{cond:3} cannot hold over the full momentum range, as they would imply vanishing baryon and quark densities. We therefore analyze the four cases separately.

\begin{enumerate}
\item $\lambda_{g \geq 0}(k) > 0\,, \ g(k) = 0$ :
    
Since $g(k) = 0$ excludes $g(k)=1$ and $\fQ(q)=1$, this case is incompatible with cases \ref{cond:2} and \ref{cond:4}, which implies $\lambda_{g \leq 1}(k) = \lambda_{\fQ \leq 1}(q) = 0$.
The condition $\fQ(q)=0$ can be compatible, allowing $\lambda_{\fQ \geq 0}(q) > 0$. From Eq.~\eqref{eq:L_var}, one finds that $\lambda_{g \geq 0}(k)$ is
\begin{equation}
    \lambda_{g\geq 0}(k) = \left[E_{\mathrm{B}}(k) - \muid - \int_{\boldsymbol{q}}  \lambda_{\fQ \geq 0}(q) \,\varphi\left(\boldsymbol{q} - \frac{\boldsymbol{k}}{N_c}\right)\right] \Theta \left[\muid - E_{\mathrm{B}}(k) \right]\,.
\end{equation}
Since $\lambda_{\fQ \geq 0}(q)$ and $\varphi\left(\boldsymbol{q} - \frac{\boldsymbol{k}}{N_c}\right)$ are positive, their convolutional integral is also positive.
Therefore, $\lambda_{g\geq 0}(k)$ never becomes positive.
This implies that $g(k) \neq 0$ for all finite ranges of momenta $k$.

\item $\lambda_{g \leq 1}(k) > 0\,, \ g(k) = 1$ :

Here $g(k)=1$ excludes cases~\ref{cond:1} and~\ref{cond:3}, so that
$\lambda_{g\ge0}(k)=\lambda_{f_Q\ge0}(q)=0$, while $f_Q(q)=1$ remains compatible allowing $\lambda_{\fQ \leq 1}(q)>0$.
From Eq.~\eqref{eq:L_var}, one finds 
\begin{equation}
    \lambda_{g\leq 1}(k)
    = \left[\muid - E_{\mathrm{B}}(k) - \int_{\boldsymbol{q}}  \lambda_{\fQ \leq 1}(q) \varphi\left(\boldsymbol{q} - \frac{\boldsymbol{k}}{N_c}\right)\right] \Theta \left[\muid - E_{\mathrm{B}}(k) \right]\,.
\end{equation}
For $\lambda_{f_Q\le1}=0$ this expression is strictly positive for
$k<\ksh$, and remains positive in part of this range for finite
$\lambda_{f_Q\le1}$. Hence regions with $g(k)=1$ are allowed.

\item $\lambda_{\fQ \geq 0}(q) > 0\,, \ \fQ(q) = 0$ :

We make use of the differential operator $\hat{L}_q$ defined in Eq.~\eqref{eq:diffOpL}. 
Since $\varphi(p)$ is the Green's function of $\hat{L}_q$, we have
\begin{equation}
    \hat{L}_q \varphi(q) = \frac{1}{\Lambda^2} (2\pi)^3 \delta^{(3)}(\bq)\,.
    \label{eq:green}
\end{equation}
By applying this to the relation $\fQ(q) = \int_{\boldsymbol{k}} \varphi(\boldsymbol{q} -\boldsymbol{k} / N_c) \fB(k)$, we find that $\fB(\Nc q) = (\Lambda^2 / \Nc^3) \hat{L}_q \fQ(q)$. Therefore, $\fQ(q) = 0$ implies $\fB(k) = 0$, and thus $g(k) = 0$, which is excluded for $k < \ksh$ by the previous analysis.

\item $\lambda_{\fQ \leq 1}(q) > 0\,, \ \fQ(q) = 1$ :

From the same relation, i.e.~$\fB(\Nc q) = (\Lambda^2 / \Nc^3) \hat{L}_q \fQ(q)$, it follows that $\fQ(q) = 1$ fixes the baryon distribution $\fB(k) = 1/\Nc^3$ implying $g(k) = 1/\Nc^3$ and vanishing multipliers $\lambda_{g \geq 0}(q) = \lambda_{g \leq 0}(q) = \lambda_{\fQ \geq 0} = 0$. 
Then Eq.~\eqref{eq:L_var} gives at $k < \ksh$ 
\begin{equation}
    E_{\mathrm{B}}(\Nc k') - \muid
    + \int_{\boldsymbol{q}}  \lambda_{\fQ \leq 1}(q) \,\varphi\left(\bq - \bk'\right) = 0\,,
\end{equation}
where $k' = k / \Nc$.
By applying $\hat{L}_{k'}$ to the formula above and using Eq.~\eqref{eq:green}, one finds
\begin{equation}
    \lambda_{\fQ \leq 1}(k / \Nc) = \muid - \Nc E_{\mathrm{Q}}(k/\Nc)\,,
\end{equation}
where $E_{\mathrm{Q}}(q)$ is the quark dispersion relation from Eq.~\eqref{eq:EQ}.
From this equation, it can be concluded that $\lambda_{\fQ \leq 1}(k / \Nc)$ is always positive for $k < \ksh$. Hence this binding case is consistent.
\end{enumerate}

The variational analysis with respect to $g(k)$ at fixed $k$ admits only two possible values, $g(k) = 1 / \Nc^3$ and $g(k) = 1$. The minimum-energy solution at fixed baryon density is consequently obtained as a piecewise function combining these two values.

The solution $g(k)=1$ for all  $k < \ksh$ violates the condition on the quark occupation $\fQ(q) < 1$ in Eq.~\eqref{eq:cond_fQ}.
Comparing all possible piecewise combinations of $1/\Nc^3$ and $1$, we find that the energy density is minimized by
\begin{equation}
\label{eq:gfactor0}
    g(k) = \frac{1}{N_c^3} \Theta(\kbu - k) + \Theta(k - \kbu)\,.
\end{equation}
This corresponds to the baryon distribution function
\begin{equation}
    \fB(k) = \frac{1}{N_c^3} \Theta(\kbu - k) + \fFD(k) \Theta(k - \kbu)\,,
    \label{eq:fB0T}
\end{equation}
where no additional factor $\Theta(\ksh - k)$ is needed in the first term as we have $\kbu\leq\ksh$ by definition.

The quark distribution for $q<q_{\mathrm{bu}}$ follows from the duality relation \eqref{eq:duality} using Eq.~\eqref{eq:fB0T}, which yields
\begin{equation}
\label{eq:fQ0T}
f_Q(q)
=
1+\frac{\sinh(q/\Lambda)}{q/\Lambda}
\Bigg[
\left(1-\frac{1}{N_c^3}\right)
e^{-q_{\mathrm{bu}}/\Lambda}\left(1+\frac{q_{\mathrm{bu}}}{\Lambda}\right)
-
e^{-q_{\mathrm{sh}}/\Lambda}\left(1+\frac{q_{\mathrm{sh}}}{\Lambda}\right)
\Bigg].
\end{equation}
For $q<q_{\mathrm{bu}}$ we require quark saturation, $f_Q(q)=1$, together with the
bound $f_Q(q)\le 1$. Since $\sinh(q/\Lambda)/(q/\Lambda)$ is positive and grows
exponentially with $q$, these conditions can only be satisfied if the bracket in
Eq.~\eqref{eq:fQ0T} vanishes, which fixes the matching momentum $\kbu$ via
\begin{equation}
\label{eq:kbu0T}
\left(1-\frac{1}{N_c^3}\right)
e^{-q_{\mathrm{bu}}/\Lambda}\left(1+\frac{q_{\mathrm{bu}}}{\Lambda}\right)
-
e^{-q_{\mathrm{sh}}/\Lambda}\left(1+\frac{q_{\mathrm{sh}}}{\Lambda}\right)
=0,
\end{equation}
with $q_{\mathrm{bu}}=k_{\mathrm{bu}}/N_c$ and $q_{\mathrm{sh}}=\ksh/N_c$.

\subsection{Determination of $g(k)$ at non-zero temperature\label{sec:densityofstatesT}}

We have shown in the previous subsection that the variational approach under inequality conditions correctly reproduces the density of states at zero temperature advocated above. We now consider the IdylliQ model at non-zero temperature and determine the density of states in this case. The procedure is analogous. We maximize the entropy density $s[g(k)]$, derived within statistical mechanics, see Eqs.~(\ref{eq:s}) and~\eqref{equ:sStatMech}, as a functional of $g(k)$ at fixed baryon density and fixed energy density. This variational problem is subject to the following constraints :
\begin{align}
    \nB[g(k)] = n^*\,, \label{eq:cond_n_finiteT}\\
    \varepsilon[g(k)] = \varepsilon^*\,, \label{eq:cond_e_finiteT}\\
    0 \leq \fFD(k) \leq 1\,, \label{eq:cond_fFD_finiteT}\\
    0 \leq g(k) \leq 1\,, \label{eq:cond_g_finiteT}\\
    0 \leq \fQ(q) \leq 1 \label{eq:cond_fQ_finiteT}\,.
\end{align}
The associated Lagrangian reads
\begin{align}
    \mathcal{L}[g(k)]
    &= s[g(k)]
    - \lambda_n (\nB[g(k)] - n^*)
    - \lambda_\varepsilon (\varepsilon[g(k)] - \varepsilon^*)\notag \\    
    &\quad + \int_{\boldsymbol{k}} \lambda_{\fFD \ge 0}(k) \fFD(k)
    - \int_{\boldsymbol{k}} \lambda_{\fFD \leq 1}(k) \left[ \fFD(k) - 1 \right]\notag \\
    &\quad + \int_{\boldsymbol{k}} \lambda_{g \ge 0}(k) g(k)
    - \int_{\boldsymbol{k}} \lambda_{g \leq 1}(k) \left[ g(k) - 1 \right]\notag \\
    &\quad + \int_{\boldsymbol{q}} \lambda_{\fQ \geq 0}(q) \fQ(q)
    - \int_{\boldsymbol{q}} \lambda_{\fQ \leq 1}(q) \left[\fQ(q) - 1\right]\,.
\end{align}
The signs of the inequality terms are opposite to those in
Eq.~\eqref{eq:lagrangian}, since the entropy density is maximized here.
From the statistical mechanical analysis we use $\fFD(k) = 1/(e^{\betaid (E_{\mathrm{B}}(k) - \muid)} + 1)$, which is the stationary point of $\mathcal{L}$ with respect to variations of
$\fFD(k)$ in the absence of the quark constraint, Eq.~\eqref{eq:cond_fQ_finiteT}.
This fixes $\lambda_\varepsilon = \betaid$ and $\lambda_n = -\betaid \muid$.

By taking the variation of $\mathcal{L}$ with respect to $g(k)$, we obtain the first KKT condition 
\begin{align}
    \label{eq:L_var_finiteT}
    0 = & \,\frac{\delta \mathcal{L}}{\delta g(k)} \\ \notag
    = & \,\left[
     \int_{\boldsymbol{q}}  \lambda_{\fQ \geq 0}(q) \,\varphi\left(\boldsymbol{q} - \frac{\boldsymbol{k}}{N_c}\right)
    - \int_{\boldsymbol{q}}  \lambda_{\fQ \leq 1}(q) \,\varphi\left(\boldsymbol{q} - \frac{\boldsymbol{k}}{N_c}\right) \right] \fFD(k) \\ \notag
    & \,- \ln[1 - \fFD(k)]
    + \lambda_{g \geq 0}(k)
    - \lambda_{g \leq 1}(k)\,.
\end{align}
The second and third KKT conditions are identical to Eqs.~(\ref{eq:KTT2g0})-(\ref{eq:KTT3fQ1}). As before, the following binding cases
may occur over connected momentum intervals in $k$ (and the
associated quark momentum $q$):
\begin{enumerate}
    \item \label{cond:1T} $\lambda_{g \geq 0}(k) > 0\,, \ g(k) = 0$ \,,
    \item \label{cond:2T} $\lambda_{g \leq 1}(k) > 0\,, \ g(k) = 1$ \,,
    \item \label{cond:3T} $\lambda_{\fQ \geq 0}(q) > 0\,, \ \fQ(q) = 0$ \,,
    \item \label{cond:4T} $\lambda_{\fQ \leq 1}(q) > 0\,, \ \fQ(q) = 1$ \,.
\end{enumerate}
Otherwise the multipliers vanish and the corresponding inequality constraints
are non-binding. We now examine which of these cases can be realized.

\begin{enumerate}
\item $\lambda_{g \geq 0}(k) > 0\,, \ g(k) = 0$ :

As at zero temperature, this case is incompatible with cases \ref{cond:2} and \ref{cond:4}, implying $\lambda_{g \leq 1}(k) = \lambda_{\fQ \leq 1}(q) = 0$. From Eq.~\eqref{eq:L_var_finiteT} one finds
\begin{equation}
    \lambda_{g\geq 0}(k) = \ln[1 - \fFD(k)] - \int_{\boldsymbol{q}}  \lambda_{\fQ \geq 0}(q) \,\varphi\left(\boldsymbol{q} - \frac{\boldsymbol{k}}{N_c}\right) \fFD(k)\,.
\end{equation}
Since $0<\fFD(k)<1$, one has $\ln[1-\fFD(k)]<0$, and since both
$\lambda_{f_Q\ge0}(q)$ and $\varphi$ are non-negative, the integral term is also
non-negative. Consequently, $\lambda_{g\ge0}(k)\le 0$ for all $k$, which contradicts the assumption $\lambda_{g\ge0}(k)>0$.
Therefore no connected momentum region with $g(k)=0$ is allowed at non-zero temperature.

\item $\lambda_{g \leq 1}(k) > 0\,, \ g(k) = 1$ :

Again, $g(k)=1$ excludes cases~\ref{cond:1} and~\ref{cond:3}, so that
$\lambda_{g\ge0}(k)=\lambda_{f_Q\ge0}(q)=0$, while $f_Q(q)=1$ remains compatible allowing $\lambda_{\fQ \leq 1}(q)>0$.
From Eq.~\eqref{eq:L_var_finiteT}, one finds that $\lambda_{g \leq 1}(k)$ is
\begin{equation}
    \lambda_{g\leq 1}(k) = -\ln[1 - \fFD(k)] - \int_{\boldsymbol{q}}  \lambda_{\fQ \leq 1}(q) \,\varphi\left(\boldsymbol{q} - \frac{\boldsymbol{k}}{N_c}\right) \fFD(k)\,.
\end{equation}
Since $0<\fFD(k)<1$, the first term is strictly positive,
$-\ln[1-\fFD(k)]>0$, while the integral term is non-negative. Consequently,
$\lambda_{g\le1}(k)$ is positive whenever the first term dominates the second.
In particular, for $\lambda_{f_Q\le1}(q)=0$ one has $\lambda_{g\le1}(k)=-\ln[1-\fFD(k)]>0$ for all $k$, showing that connected momentum regions with $g(k)=1$ are allowed at non-zero temperature.

\item $\lambda_{\fQ \geq 0}(q) > 0\,, \ \fQ(q) = 0$ :

Using the relation $\fB(\Nc q) = (\Lambda^2 / \Nc^3) \hat{L}_q \fQ(q)$ 
one finds that $f_Q(q)=0$ implies $f_B(k)=0$, and therefore $g(k)=0$.
As shown above, no connected momentum region with $g(k)=0$ is allowed at non-zero
temperature. Hence this binding case cannot be realized.

\item $\lambda_{\fQ \leq 1}(q) > 0\,, \ \fQ(q) = 1$ :

Using $f_B(N_c q)=\frac{\Lambda^2}{N_c^3}\hat L_q f_Q(q)$,
one finds that $f_Q(q)=1$ implies $f_B(k)=\frac{1}{N_c^3}$. Since $f_B(k)=g(k)\fFD(k)$, this fixes $g(k)=\frac{1}{N_c^3\fFD(k)}$, and therefore $\lambda_{g\ge0}(k)=\lambda_{g\le1}(k)=\lambda_{f_Q\ge0}(q)=0$.

Equation~\eqref{eq:L_var_finiteT} then yields the condition
\begin{equation}
-\ln[1-\fFD(N_c k')]
-
\int_{\boldsymbol q}\lambda_{f_Q\le1}(q)
\varphi(\boldsymbol q-\boldsymbol k')\fFD(N_c k')
=0,
\end{equation}
where we introduced $k'=k/N_c$.
Applying $\hat L_{k'}$ and using Eq.~\eqref{eq:green}, one obtains
\begin{equation}
\lambda_{f_Q\le1}(k')
=
-\Lambda^2 \hat L_{k'}
\left[
\frac{1}{\fFD(N_c k')}
\ln\!\bigl(1-\fFD(N_c k')\bigr)
\right].
\end{equation}
The right-hand side is strictly positive for all $k'$, and therefore
$\lambda_{f_Q\le1}(k')>0$, showing that this binding case is consistently
realized at finite temperature.
\end{enumerate}

These considerations, based on the variational analysis with respect to $g(k)$ at
fixed $k$, leave us again with two possibilities,
$g(k)=1/[N_c^3\fFD(k)]$ and $g(k)=1$. In analogy to the zero-temperature case
discussed above, the maximum entropy at fixed $\nB$ and $\varepsilon$ is obtained
for a piecewise function,
\begin{equation}
    \label{eq:gfactor}
    g(k) = \frac{1}{N_c^3 \fFD(k)} \Theta(\kbu - k) + \Theta(k - \kbu)\,.
\end{equation}
According to Eq.~\eqref{eq:gfFD}, this yields the baryon distribution
\begin{equation}
    \fB(k) = \frac{1}{N_c^3} \Theta(\kbu - k) + \fFD(k)\,\Theta(k - \kbu)\,.
    \label{eq:fBT}
\end{equation}
As one can see, the low-momentum part of $\fB(k)$ up to $\kbu$ does not depend on the Fermi-Dirac distribution. It is fully determined by the quark saturation constraint together with the specific kernel of the duality relation. As we shall see in section~\ref{sec:6}, this part has strong consequences on the physical temperature of the system. We note that the form of $g(k)$ in Eq.~\eqref{eq:gfactor} reduces to the expression in Eq.~\eqref{eq:gfactor0} at $T=0$ knowing that in this limit $\fFD(k)=1$ for $k\leq\kbu$.

The switching momentum $k_{\mathrm{bu}}$, separating the suppressed core from the thermal outer shell of the baryon distribution function can be found from the onset condition of quark saturation. The quark distribution function $f_Q(q)$ for quark momentum $q=k/N_c$ is obtained from Eq.~\eqref{eq:duality} using Eq.~\eqref{eq:fBT}. In the small momentum region $q<\qbu=\kbu/N_c$ one finds 
\begin{equation}
\label{eq:fQsmall}
    \fQ(q) = 1 + \frac{\sinh(q/\Lambda)}{q/\Lambda} \left[\frac{\Nc^3}{\Lambda^2}\int_{\qbu}^\infty dp\, p\, e^{-p/\Lambda} \fFD(N_cp) - e^{-\qbu/\Lambda}\left(1 + \frac{\qbu}{\Lambda}\right) \right]\,.
\end{equation}
In the $\Tid\to 0$ limit this equation reduces to the one given in Eq.~\eqref{eq:fQ0T}. Since $\sinh(q/\Lambda)/(q/\Lambda)>0$, Eq.~\eqref{eq:fQsmall} shows that the bound
$f_Q(q)\le 1$ for $0\le q\le q_{\mathrm{bu}}$ is equivalent to requiring the
bracket to be non-positive. Imposing quark saturation, $f_Q(q)=1$ for
$q\le q_{\mathrm{bu}}$, fixes the bracket to zero and yields
\begin{equation}
    \label{eq:kbuT}
    \frac{\Nc^3}{\Lambda^2}\int_{\qbu}^\infty dp\, p\, e^{-p/\Lambda} \fFD(\Nc p) - e^{-\qbu/\Lambda}\left(1 + \frac{\qbu}{\Lambda}\right) = 0 \, ,
\end{equation}
which determines $\qbu$ (and thus $\kbu$) for given $\Tid$ and $\muid$. In the $\Tid\to 0$ limit this equation becomes Eq.~\eqref{eq:kbu0T}. Equation~\eqref{eq:kbuT} implies that if $\qbu=0$ for a given $\Tid$ and $\muid$ there is no quark saturation for any $q$ and we are in the normal baryonic phase.

We note that in the opposite momentum region, $q>\qbu$, one finds 
\begin{align}
    \fQ(q)
    &= \left[\frac{\qbu}{\Lambda} \cosh \left(\frac{\qbu}{\Lambda}\right) - \sinh \left(\frac{\qbu}{\Lambda}\right)\right] \frac{e^{-q/\Lambda}}{q/\Lambda}\notag \\
    &\quad+ \frac{\Nc^3}{\Lambda^2} \frac{e^{-q/\Lambda}}{q/\Lambda} \int_{\qbu}^{q} dp\, p\, \sinh \left(\frac{p}{\Lambda}\right) \fFD(\Nc p)\notag\\
    &\quad+ \frac{\Nc^3}{\Lambda^2} \frac{\sinh(q/\Lambda)}{q/\Lambda} \int_{q}^\infty dp\,p\, e^{-p/ \Lambda} \fFD(\Nc p)\,.
    \label{eq:fQ}
\end{align}
Although the factor $\sinh(q/\Lambda)$ grows exponentially, the integral in the
last term decays as $\sim e^{-q/\Lambda}$ due to its lower limit at $p=q$, so that
the exponential behaviour cancels and $\fQ(q)$ remains finite for large $q$. One can verify from Eqs.~\eqref{eq:fQsmall} and~\eqref{eq:fQ} that $\fQ(q)$ is a continuous function at $\qbu$.

\begin{figure}
    \centering
    \includegraphics[width=0.9995\linewidth]{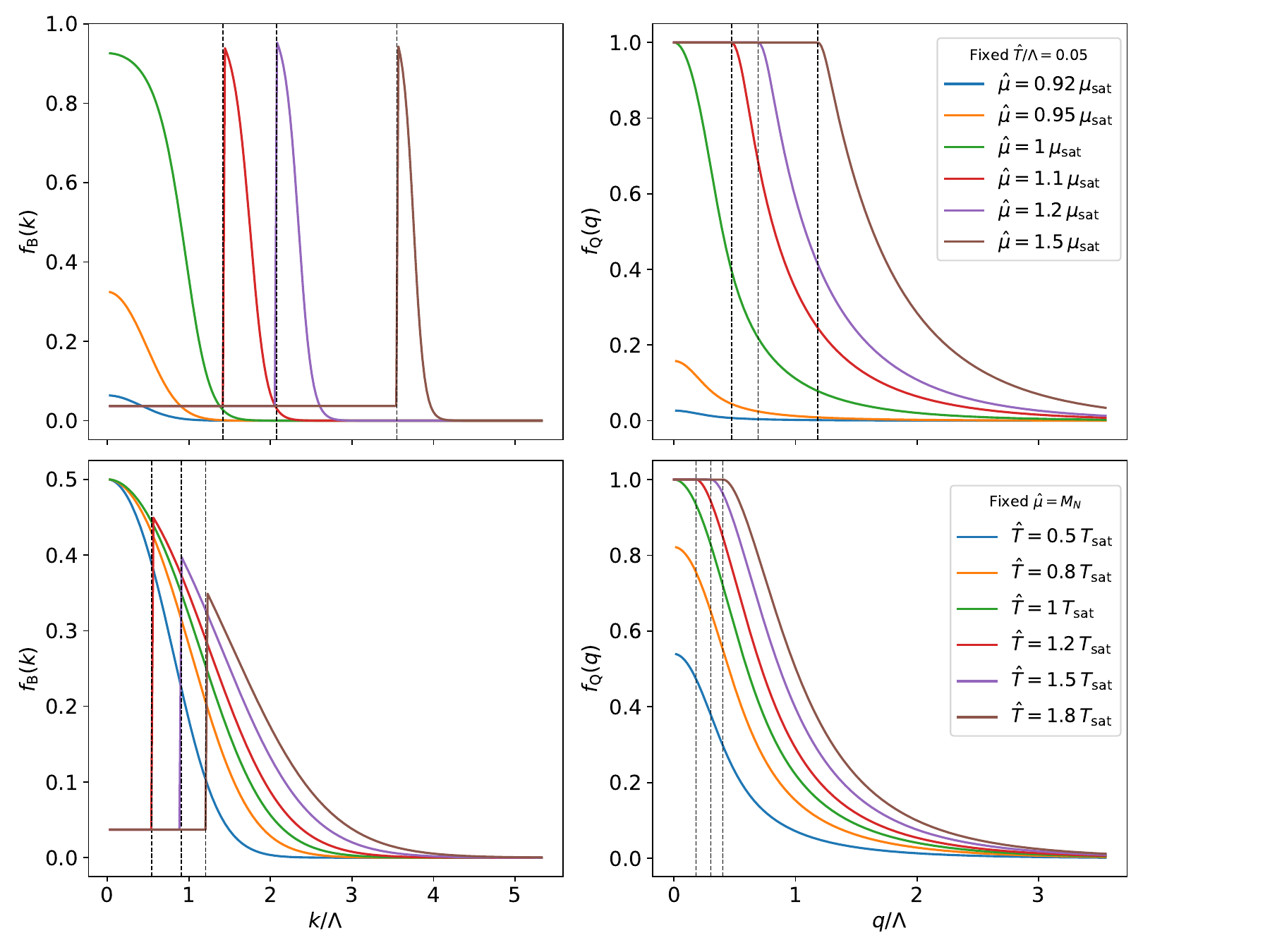}
    \caption{The baryon distribution function $\fB(k)$ (left) as a function of baryon momentum $k$  and the corresponding quark distribution function per color degree of freedom $\fQ(q)$ (right) as a function of quark momentum $q$ at a fixed $\Tid=0.01$~GeV (upper panels) and at a fixed $\muid=M_N$ (lower panels). The dashed vertical lines indicate the position of the bulk momentum $\kbu$ and $\qbu$ in the Quarkyonic Matter regime.
    With $\Lambda = 0.3\,\text{GeV}$ and $M_N = 0.94\,\text{GeV}$, we find $\mu_{\rm sat} = 0.9784$~GeV and $T_{\rm sat} = 0.0621$~GeV.}
    \label{fig:fBandfQ}
\end{figure}

In Fig.~\ref{fig:fBandfQ} we show the baryon (left panels) and quark (right
panels) distribution functions for different values of $\Tid$ and $\muid$. At
fixed $\Tid$ (top row), increasing $\muid$ drives the quark distribution towards
saturation at small momentum. Above a threshold value $\mu_{\mathrm{sat}}$ one
finds $f_Q(q)=1$ for $0\le q\le q_{\mathrm{bu}}$, signalling the onset of the Quarkyonic Matter regime. In the baryon distribution this is reflected by the emergence
of a low-momentum core with $f_B(k)=1/N_c^3$ for $0\le k\le k_{\mathrm{bu}}$,
followed by an outer shell governed by the Fermi--Dirac form. 
At fixed $\muid$ (bottom row), increasing $\Tid$ broadens the thermal shell and
can likewise induce a saturated region in $f_Q(q)$ for $q\leq q_{\mathrm{bu}}$ above
a characteristic temperature $T_{\mathrm{sat}}$.

\begin{figure}
    \centering
    \includegraphics[width=0.6\linewidth]{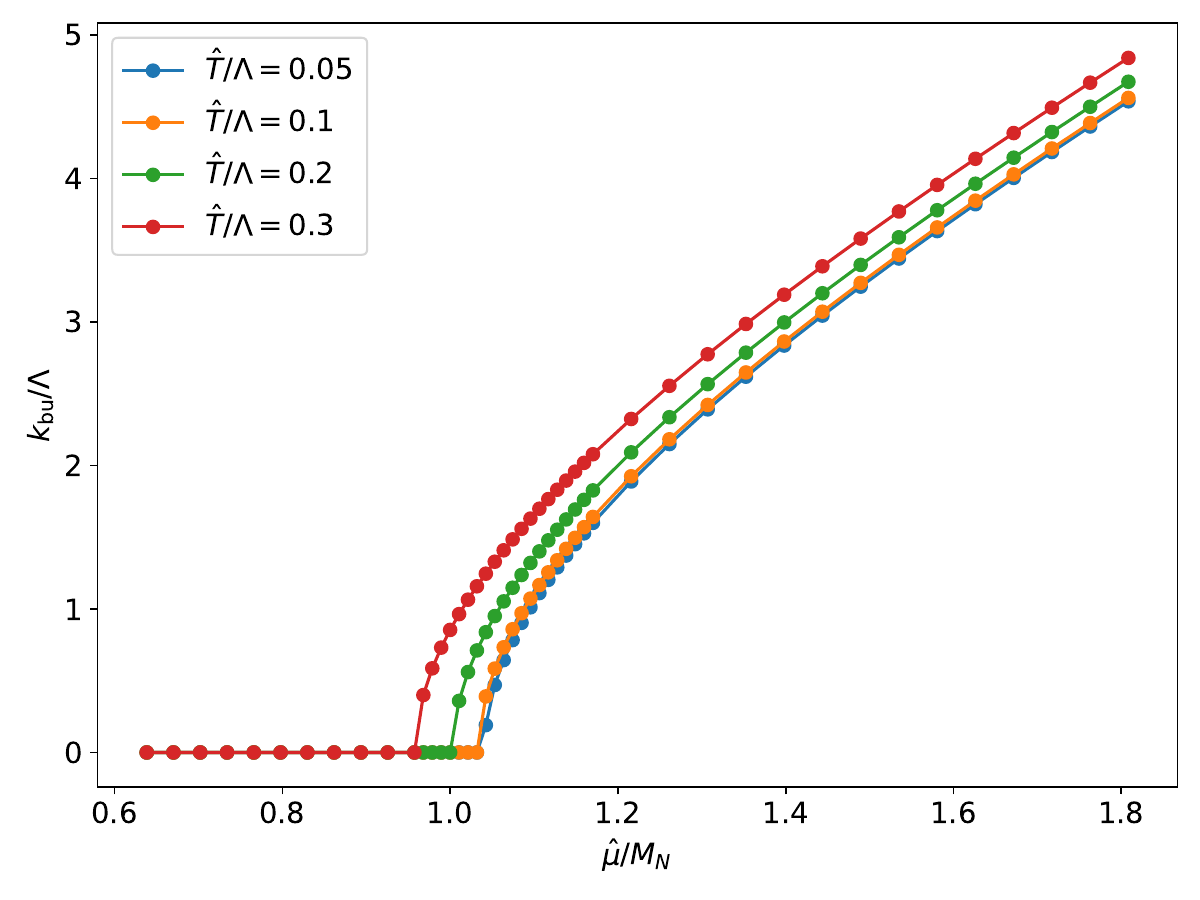}
    \caption{Scaled bulk momentum $\kbu/\Lambda$ as a function of $\muid/M_N$ for different values of $\Tid/\Lambda$.}
    \label{fig:kbu}
\end{figure}

In Fig.~\ref{fig:kbu}, the behavior of the bulk momentum $\kbu$ is shown as a function of $\muid/M_N$ for different values of $\Tid/\Lambda$. As long as $\kbu=0$ for given $\Tid$ and $\muid$, the quark distribution is not saturated, which corresponds to the normal baryonic phase. For fixed $\Tid$, increasing $\muid$ leads to a nonzero $\kbu$ at a threshold value $\mu_{\rm sat}$, signaling
the onset of the Quarkyonic Matter regime. The saturation chemical potential $\mu_{\rm sat}$ decreases with increasing $\Tid$, analogous to the behavior of the saturation curve in the QCD phase diagram discussed in Ref.~\cite{Bluhm:2024uhj}. Moreover, for fixed $\muid$ within the Quarkyonic Matter phase, the value of $\kbu$ increases with increasing $\Tid$, consistent with the behavior of the baryon distribution function $\fB(k)$ observed in Fig.~\ref{fig:fBandfQ}.

\section{Physical chemical potential and temperature \label{sec:6}}

\begin{figure}[t]
    \centering
    \includegraphics[width=0.99\linewidth]{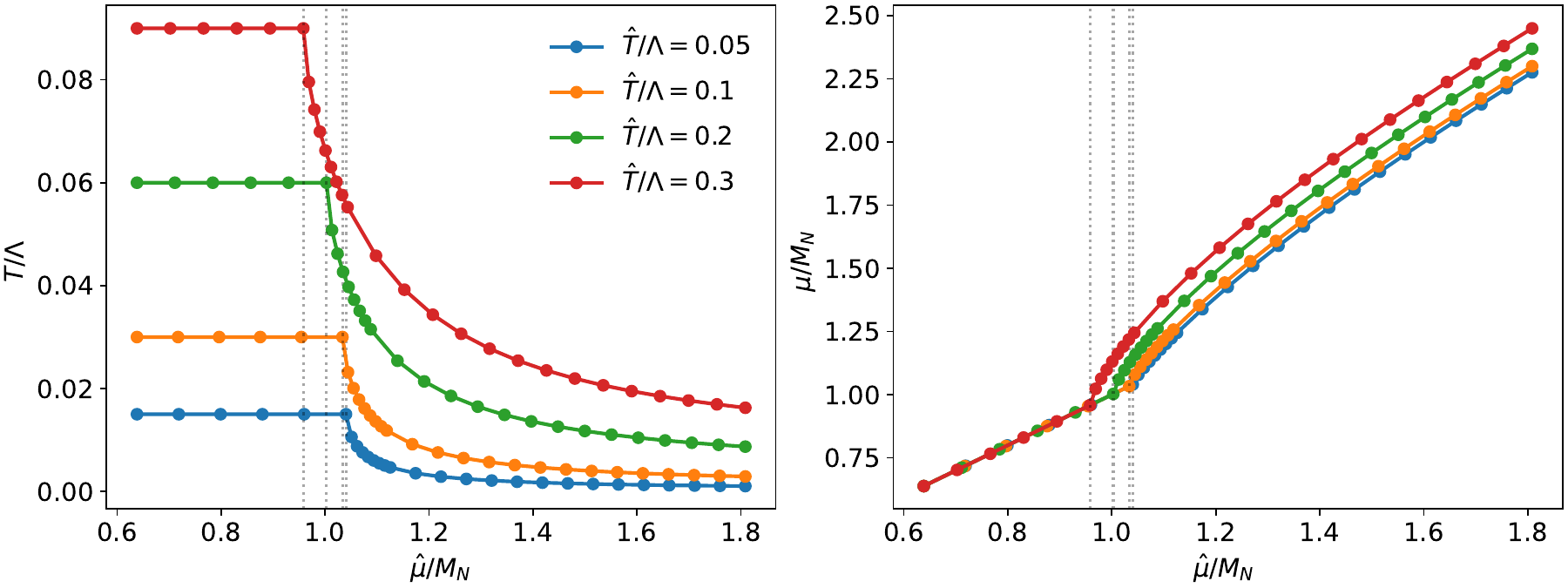}
    \caption{The scaled physical temperature $T/\Lambda$ (left panel) and physical baryon chemical potential $\muB/\Lambda$ (right panel) as functions of the scaled Lagrange multiplier $\muid/M_N$ for different values of $\Tid/\Lambda$.}
    \label{fig:Tphys}
\end{figure}

So far we have expressed the thermodynamic quantities $\nB$, $\varepsilon$, and $s$ as functions of the Lagrange multipliers $\Tid$ and $\muid$. For physical applications, such as the construction of an equation of state, these quantities must however be expressed in terms of the physical temperature $T$ and baryon
chemical potential $\mu_B$. The latter are obtained from the thermodynamic relations 
\begin{align}
\label{eq:physT}
    T &= \left(\frac{\partial \varepsilon}{\partial s}\right)_{\nB}
    = \frac{\partial (\varepsilon, \nB)}{\partial (s, \nB)}
    = \frac{\frac{\partial (\varepsilon, \nB)}{\partial (\Tid, \muid)}}{\frac{\partial (s, \nB)}{\partial (\Tid, \muid)}}
    = \frac{\left(\frac{\partial \varepsilon}{\partial \Tid}\right)_{\muid} \left(\frac{\partial \nB}{\partial \muid}\right)_{\Tid} - \left(\frac{\partial \nB}{\partial \Tid}\right)_{\muid} \left(\frac{\partial \varepsilon}{\partial \muid}\right)_{\Tid}}
    {\left(\frac{\partial s}{\partial \Tid}\right)_{\muid} \left(\frac{\partial \nB}{\partial \muid}\right)_{\Tid} - \left(\frac{\partial \nB}{\partial \Tid}\right)_{\muid} \left(\frac{\partial s}{\partial \muid}\right)_{\Tid}}\,,\\
    \label{eq:physmu}
    \mu_B &= \left(\frac{\partial \varepsilon}{\partial \nB}\right)_s
    = \frac{\partial (\varepsilon, s)}{\partial (\nB, s)}
    = \frac{\frac{\partial (\varepsilon, s)}{\partial (\Tid, \muid)}}{\frac{\partial (\nB, s)}{\partial (\Tid, \muid)}}
    = \frac{\left(\frac{\partial s}{\partial \Tid}\right)_{\muid} \left(\frac{\partial \varepsilon}{\partial \muid}\right)_{\Tid}-\left(\frac{\partial \varepsilon}{\partial \Tid}\right)_{\muid} \left(\frac{\partial s}{\partial \muid}\right)_{\Tid}}
    {\left(\frac{\partial s}{\partial \Tid}\right)_{\muid} \left(\frac{\partial \nB}{\partial \muid}\right)_{\Tid}-\left(\frac{\partial \nB}{\partial \Tid}\right)_{\muid} \left(\frac{\partial s}{\partial \muid}\right)_{\Tid}}\,.
\end{align}
Detailed expressions of the derivatives are given in Appendix~\ref{sec:derivs}. As long as the system remains in the normal baryonic phase with $\kbu=0$, one finds $T=\Tid$ and $\mu_B=\muid$. Once the system enters the Quarkyonic Matter phase, the physical temperature and baryon chemical potential deviate from the Lagrange multipliers due to the presence of the low-momentum saturated core in the baryon distribution function up to
$k_{\mathrm{bu}}$. In Fig.~\ref{fig:Tphys} we show numerical results for $T$ and $\mu_B$ as functions of $\muid/M_N$ for different values of $\Tid$. In the Quarkyonic Matter phase the
physical temperature is significantly reduced compared to $\Tid$, while the physical baryon chemical potential exceeds $\muid$. Moreover, the onset of these deviations shifts to lower values of $\muid$ with increasing $\Tid$, in agreement with the behavior of the saturation line in the QCD phase diagram discussed in
Ref.~\cite{Bluhm:2024uhj}. In the limit $\Tid\to0$, where the entropy density vanishes, Eq.~\eqref{eq:physmu}
reduces to $\mu_B=\partial\varepsilon/\partial\nB$, while Eq.~\eqref{eq:physT} yields $T=0$, as expected.

\section{Summary and conclusions}

In the IdylliQ model framework, the existence of Quarkyonic Matter can be understood as the result of simultaneous phase-space occupation constraints acting on quarks and on the baryons that contain them, originating from the Pauli exclusion principle. As established in an earlier work at zero temperature~\cite{Fujimoto:2023mzy}, the baryon distribution function $\fB$ exhibits a shell-like structure, in which the occupation probability for momenta below the density-dependent bulk momentum $\kbu$ is suppressed by a factor $1/\Nc^3$ in comparison to an ideal Fermi gas of nucleons. An immediate consequence of this suppression is that the energy of the highest occupied energy state no longer coincides with the baryon chemical potential $\muB$. 

The central goal of the present work is to extend this description of Quarkyonic Matter to non-zero temperatures $T$. A direct application of standard finite-temperature expressions, however, immediately encounters a fundamental problem. Due to the suppression of the baryon distribution at low momenta in the Quarkyonic Matter phase, the standard textbook expression Eq.~\eqref{eq:snaiv} for the entropy density of an ideal Fermi gas does not vanish in the zero-temperature limit, in apparent violation
of the third law of thermodynamics. We therefore need to consistently reformulate the entropy density in Quarkyonic Matter in the presence of the additional phase-space constraint for the quark occupation. 

We resolve this problem by formulating the thermodynamics of Quarkyonic Matter in terms of a restricted Fock space of physically allowed baryonic states. The restriction arises from the Pauli exclusion principle acting on the quark degrees of freedom inside different baryons, which reduces the number of available baryon states in a momentum-dependent manner once quark saturation is
reached. This reduction is encoded in a momentum-dependent density of states $g(k)$, such that the baryon distribution function factorizes as $f_B(k)=g(k)f_{\rm FD}(k)$. With this separation between thermal occupation and state counting, the entropy density is consistently defined and vanishes in the zero-temperature limit, while reproducing the known shell structure of Quarkyonic Matter.

The resulting formulation of the entropy density is derived from first principles using methods of quantum statistical mechanics which we adapted to systems with a restricted Fock space. Starting from Boltzmann’s definition of the entropy, we construct the grand canonical ensemble for a system in which individual one-particle states are either allowed or forbidden due to the quark saturation constraints. This constructions follows the method of Darwin and Fowler, which provides a framework to implement additional constraints naturally within a generating functional. This  generating functional is evaluated in the saddle-point approximation, leading to a consistent expression for the entropy density in the presence of a restricted Fock space. In order to determine the explicit form of the momentum-dependent density of states, we employ a variational approach with inequality constraints, implemented via the Karush–Kuhn–Tucker conditions. This procedure uniquely determines the density of states both at zero temperature, where it follows from minimizing the energy density, and at non-zero temperature, where it follows from maximizing the entropy density.

The principle result of this work is the emergence of a momentum-dependent density of states for baryons in Quarkyonic Matter. Once quark saturation sets in, the baryonic phase space separates into two distinct regions: a low-momentum bulk region in which the number of physically allowed baryon states is strongly reduced due to quark Pauli blocking, and an outer shell region where the
density of states is that of an ideal Fermi gas. While the bulk region appears underoccupied when viewed solely through the baryon distribution function, it is in fact fully occupied once the reduced density of states is taken into account.

Similarly as at zero temperature, a direct consequence of this modified density of states is that the thermodynamic Lagrange multipliers $\Tid$ and $\muid$ entering the Fermi--Dirac distribution no longer coincide with the physical temperature $T$ and baryon chemical potential $\mu_B$ in the Quarkyonic Matter phase. Because the saturated bulk region does not participate in thermal excitations, the entropy associated with a small change in energy can be significant, leading to a physical temperature that is reduced compared to $\Tid$, while the physical baryon chemical potential exceeds $\muid$.

With the thermodynamics of Quarkyonic Matter now formulated in a consistent and well-defined way, this framework enables us to systematically determine further thermodynamic properties at non-zero temperature. In a forthcoming publication we will use the results established here to compute the pressure, the equation of state, and related response functions such as susceptibilities.
These quantities are essential inputs for exploratory studies of Quarkyonic Matter in hydrodynamic and transport approaches, where transport coefficients will also play a central role. In particular, the equation of state of Quarkyonic Matter at small but non-zero temperature provides a promising basis for investigating its impact on the structure and evolution of dense astrophysical systems, including young and even hotter proto neutron stars as well as neutron star mergers.

\begin{acknowledgments}
    The authors thank Larry D. McLerran for many stimulating discussions and his continuous support. Y.F. thanks Toru~Kojo and Dam~Thanh~Son for insightful conversations. M.B. and M.N. thank the Institute for Nuclear Theory at the University of Washington for its kind hospitality and stimulating research environment. This research was supported in part by the INT's U.S. Department of Energy grant No. DE-FG02-00ER41132. M.N. acknowledges support from the long term visitor program at the INT.
\end{acknowledgments}

\appendix

\section{The method of the most probable distribution}
\label{sec:dirac}

We derive the Fermi-Dirac distribution following the approach taken in the original derivation by Dirac~\cite{Dirac:1926jz}.
As we will mention below, this approach does not directly reflect the modification in the density of states due to the Quarkyonic nature of the matter, which is captured by the factor $g_i$.
Let us divide all the quantum states into groups each of which contains states with nearly equal energy.
We call each group a cell and label it by the Greek index $\sigma$.
Let the energy of the $\sigma$-th cell be $E_\sigma$ and the number of states contained in it be $A_\sigma$.
A state of the whole system can then be specified by the set $\bN = (N_1, N_2, \ldots, N_\sigma, \ldots)$ of numbers of particles in each cell. To find the statistical weight $W(\bN)$ for a Fermi gas, it is sufficient to find $w_\sigma$, the number of ways in which $N_\sigma$ particles can be assigned to the $\sigma$-th cell, which contains $A_\sigma$ energy levels. Since interchanging particles from different cells does not lead to a new state of the system, we have $W(\bN) = \prod_\sigma w_\sigma$.

The number of particles occupying any of the $A_\sigma$ levels in the $\sigma$-th cell is either 0 or 1. Therefore $w_\sigma$ is equal to the number of ways in which $N_\sigma$ can be chosen from $A_\sigma$, i.e.
\begin{equation}
    w_\sigma
    = \binom{A_\sigma}{N_\sigma}
    = \frac{A_\sigma!}{N_\sigma! (A_\sigma - N_\sigma)!}\,.
\end{equation}
Then,
\begin{equation}
    W(\bN) = \prod_\sigma w_\sigma = \prod_\sigma \frac{A_\sigma!}{N_\sigma! (A_\sigma - N_\sigma)!}\,.
\end{equation}

To obtain the entropy from Boltzmann's formula Eq.~\eqref{eq:Boltzmann}, we need to sum $W(\bN)$ over $\bN$ under the conditions 
\begin{equation}
    E = \sum_\sigma E_\sigma N_\sigma\,,\qquad
    N = \sum_\sigma N_\sigma\,.
\end{equation}
However, this is a formidable task.
So, instead, we approximate $\mathcal{W}$ by the maximum statistical weight $W(\bN^*)$, where $\bN^*$ is the set of occupation numbers that maximizes $W(\bN)$ subject to the conditions above.
In the micro canonical distribution, the distribution that maximizes $W(\bN)$ subject to the conditions above is the most probable one.
To obtain $\bN^*$, we take the logarithm of $W(\bN)$ and put the variation with respect to $\bN$ equal to zero.
Since the total energy and number of particles are fixed, the conditions are
\begin{equation}
\begin{split}
    \delta \ln W(\bN) &= \sum_\sigma \ln \left(\frac{A_\sigma}{N_\sigma} - 1\right) \delta N_\sigma =0 \,,\\
    \delta E &= \sum_\sigma E_\sigma \delta N_\sigma = 0\,,\\
    \delta N &= \sum_\sigma \delta N_\sigma = 0\,,
\end{split}
\end{equation}
where we have used Stirling's formula $\ln N! \approx N \ln N - N$ in $\ln W$. By applying the method of Lagrange multipliers, we have
\begin{equation}
    \delta \ln W(\bN) - \alpha \delta N - \beta \delta E 
    = \sum_\sigma \left[\ln \left(\frac{A_\sigma}{N_\sigma} - 1\right) - \alpha - \beta E_\sigma \right] \delta N_\sigma 
    =0\,,
\end{equation}
where $\alpha$ and $\beta$ are Lagrange multipliers, which can be determined by the thermodynamic relations.
Hence, we get the most probable distribution as
\begin{equation}
    n_\sigma^* \equiv \frac{N_\sigma^*}{A_\sigma} = \frac{1}{e^{\beta E_\sigma + \alpha} + 1}\,.
\end{equation}
The entropy is obtained from Boltzmann's formula
\begin{align}
    S
    \simeq \ln W(\bN^*)
    &=  \sum_\sigma \ln \frac{A_\sigma!}{N_\sigma^*! (A_\sigma - N_\sigma^*)!} \notag \\
    &\simeq \sum_\sigma \left[\left(N_\sigma^* - A_\sigma \right) \ln \left(\frac{A_\sigma}{N_\sigma^*} - 1\right) + A_\sigma \ln \frac{A_\sigma}{N_\sigma^*} \right] \notag \\
    &= - \sum_\sigma A_\sigma \left[n_\sigma^* \ln n_\sigma^* + (1 - n_\sigma^*) \ln (1 - n_\sigma^*) \right]\,,
\end{align}
where we have used Stirling's formula again.

There is a problem with this derivation when the quark Pauli exclusion principle is taken into account as additional constraint. It is unclear how the quark Pauli exclusion principle in the dual Quarkyonic picture affects the final expression of the entropy. So far, we have grouped the quantum states into cells comprising a substantial number of microscopic states in order to apply Stirling's formula, while the modification due to the quark Pauli exclusion can only be quantified by the factor $g_i$ which appears in each sublevel of the cell. It is thus ambiguous how this factor should be handled in the grouping of the quantum states into cells. Therefore, we chose to give a derivation in Sec.~\ref{sec:derivation} that does not rely on the grouping of states.

\section{Derivative expressions for the physical temperature and baryon chemical potential}
\label{sec:derivs}

We summarize the derivative expressions appearing in the definitions of the physical temperature $T$ and the physical baryon chemical potential $\muB$ in equations~\eqref{eq:physT} and~\eqref{eq:physmu}. Based on the expressions for the baryon-, energy- and entropy densities in Eqs.~\eqref{eq:n},~\eqref{eq:e} and~\eqref{eq:s} together with Eqs.~\eqref{eq:gfactor} and~\eqref{eq:fBT}, the thermodynamic quantities depend both explicitly on $\Tid$ and $\muid$ as well as implicitly. The implicit dependence is a consequence of the $\Tid$- and $\muid$-dependence of $\kbu=\Nc\,\qbu$ according to Eq.~\eqref{eq:kbuT}. This latter equation allows us to determine the necessary derivatives of $\kbu$ with respect to $\Tid$ and $\muid$. 

We start with the expression for the baryon density based on Eq.~\eqref{eq:n} for $d$ degenerate degrees of freedom 
\begin{equation}
    n_B = \frac{d}{2\pi^2N_c^3} \int_0^{\kbu} dk\,k^2 + \frac{d}{2\pi^2} \int_{\kbu}^\infty dk\,k^2 \fFD(k)\,.
\end{equation} 
In Eqs.~\eqref{eq:physT} and~\eqref{eq:physmu} we need to calculate
\begin{equation}
     \left(\frac{\partial \nB}{\partial \Tid}\right)_{\muid} = \frac{\partial \nB}{\partial \Tid_{\rm ex}} + \frac{\partial \nB}{\partial{\kbu}} \left(\frac{\partial \kbu}{\partial{\Tid}}\right)_{\muid} \,,
\end{equation}
where 
\begin{eqnarray}
    \frac{\partial \nB}{\partial \Tid_{\rm ex}} & = & \frac{d}{2\pi^2} \int_\kbu^\infty dk\,k^2 \frac{(E(k)-\muid)}{\Tid^2} \fFD(k)(1-\fFD(k)) \,, \\
    \frac{\partial \nB}{\partial{\kbu}} & = & \frac{d}{2\pi^2N_c^3} \kbu^2 - \frac{d}{2\pi^2} \kbu^2 \fFD(\kbu)
\end{eqnarray}
and, following Eq.~\eqref{eq:kbuT}, 
\begin{equation}
    \left(\frac{\partial \kbu}{\partial{\Tid}}\right)_{\muid} = \frac{N_c^3\,e^{\kbu/(\Lambda N_c)}}{\kbu(N_c^3\fFD(\kbu)-1)} \int_\kbu^\infty dk\,k\,e^{-k/(\Lambda N_c)} \frac{(E(k)-\muid)}{\Tid^2} \fFD(k)(1-\fFD(k)) \,.
\end{equation}
Similarly, we find 
\begin{equation}
     \left(\frac{\partial \nB}{\partial \muid}\right)_{\Tid} = \frac{\partial \nB}{\partial \muid_{\rm ex}} + \frac{\partial \nB}{\partial{\kbu}} \left(\frac{\partial \kbu}{\partial{\muid}}\right)_{\Tid}
\end{equation}
with 
\begin{equation}
    \frac{\partial \nB}{\partial \muid_{\rm ex}} = \frac{d}{2\pi^2} \int_\kbu^\infty dk\,k^2 \frac{1}{\Tid} \fFD(k)(1-\fFD(k))
\end{equation}
and 
\begin{equation}
    \left(\frac{\partial \kbu}{\partial{\muid}}\right)_{\Tid} = \frac{N_c^3\,e^{\kbu/(\Lambda N_c)}}{\kbu(N_c^3\fFD(\kbu)-1)} \int_\kbu^\infty dk\,ke^{-k/(\Lambda N_c)} \frac{1}{\Tid} \fFD(k)(1-\fFD(k))
\end{equation}
from using Eq.~\eqref{eq:kbuT}.

The other derivative expressions follow from similar calculations. For the derivatives of the energy density in Eq.~\eqref{eq:e} with degeneracy factor $d$
\begin{equation}
    \varepsilon = \frac{d}{2\pi^2N_c^3} \int_0^{\kbu} dk\,k^2 E(k) + \frac{d}{2\pi^2} \int_{\kbu}^\infty dk\,k^2 E(k) \fFD(k)
\end{equation}
we find 
\begin{eqnarray}
     \left(\frac{\partial\varepsilon}{\partial \Tid}\right)_{\muid} & = & \frac{\partial\varepsilon}{\partial \Tid_{\rm ex}} + \frac{\partial\varepsilon}{\partial{\kbu}} \left(\frac{\partial \kbu}{\partial{\Tid}}\right)_{\muid} \,, \\
     \left(\frac{\partial\varepsilon}{\partial \muid}\right)_{\Tid} & = & \frac{\partial\varepsilon}{\partial \muid_{\rm ex}} + \frac{\partial\varepsilon}{\partial{\kbu}} \left(\frac{\partial \kbu}{\partial{\muid}}\right)_{\Tid}
\end{eqnarray}
with 
\begin{eqnarray}
    \frac{\partial\varepsilon}{\partial \Tid_{\rm ex}} & = & \frac{d}{2\pi^2} \int_\kbu^\infty dk\,k^2 E(k) \frac{(E(k)-\muid)}{\Tid^2} \fFD(k)(1-\fFD(k)) \,, \\
    \frac{\partial\varepsilon}{\partial \muid_{\rm ex}} & = & \frac{d}{2\pi^2} \int_\kbu^\infty dk\,k^2 \frac{E(k)}{\Tid} \fFD(k)(1-\fFD(k))
\end{eqnarray}
and 
\begin{equation}
    \frac{\partial\varepsilon}{\partial{\kbu}} = \frac{d}{2\pi^2N_c^3} \kbu^2 E(\kbu) - \frac{d}{2\pi^2} \kbu^2 E(\kbu) \fFD(\kbu) \,.
\end{equation}
For the derivatives of the entropy density in Eq.~\eqref{eq:s} reading explicitly
\begin{align}
    \nonumber
    s = & -\frac{d}{2\pi^2N_c^3} \int_0^{\kbu} dk\,k^2 \left[\ln\fFD(k)+\left(\frac{1}{\fFD(k)}-1\right)\ln(1-\fFD(k))\right] \\ 
    & - \frac{d}{2\pi^2} \int_{\kbu}^\infty dk\,k^2 \left[\fFD(k)\ln\fFD(k)+\left(1-\fFD(k)\right)\ln(1-\fFD(k))\right]
\end{align}
we find 
\begin{eqnarray}
     \left(\frac{\partial s}{\partial \Tid}\right)_{\muid} & = & \frac{\partial s}{\partial \Tid_{\rm ex}} + \frac{\partial s}{\partial{\kbu}} \left(\frac{\partial \kbu}{\partial{\Tid}}\right)_{\muid} \,, \\
     \left(\frac{\partial s}{\partial \muid}\right)_{\Tid} & = & \frac{\partial s}{\partial \muid_{\rm ex}} + \frac{\partial s}{\partial{\kbu}} \left(\frac{\partial \kbu}{\partial{\muid}}\right)_{\Tid}
\end{eqnarray}
with 
\begin{eqnarray}
    \nonumber
    \frac{\partial s}{\partial \Tid_{\rm ex}} & = & -\frac{d}{2\pi^2 N_c^3} \int_0^\kbu dk\,k^2 \frac{(E(k)-\muid)}{\Tid^2} \left(1-\frac{1}{\fFD(k)}\right)\ln(1-\fFD(k)) \\
    & & + \frac{d}{2\pi^2} \int_\kbu^\infty dk\,k^2 \frac{(E(k)-\muid)^2}{\Tid^3} \fFD(k)(1-\fFD(k)) \,, \\
    \nonumber
    \frac{\partial s}{\partial \muid_{\rm ex}} & = & -\frac{d}{2\pi^2 N_c^3} \int_0^\kbu dk\,k^2 \frac{1}{\Tid} \left(1-\frac{1}{\fFD(k)}\right)\ln(1-\fFD(k)) \\
    & & + \frac{d}{2\pi^2} \int_\kbu^\infty dk\,k^2 \frac{(E(k)-\muid)}{\Tid^2} \fFD(k)(1-\fFD(k)) \,.
\end{eqnarray}
Finally, we have 
\begin{align}
    \nonumber
    \frac{\partial s}{\partial{\kbu}} = &  -\frac{d}{2\pi^2 N_c^3} \kbu^2 \left[\ln\fFD(\kbu)+\left(\frac{1}{\fFD(\kbu)}-1\right)\ln(1-\fFD(\kbu))\right]  \\
    & + \frac{d}{2\pi^2} \kbu^2 \left[\fFD(\kbu)\ln\fFD(\kbu)+\left(1-\fFD(\kbu)\right)\ln(1-\fFD(\kbu))\right] \,.
\end{align}

In the normal baryonic phase, when $\kbu=0$, the above implicit derivatives all vanish and we are left with 
\begin{eqnarray}
    T=\frac{{\cal A}}{{\cal A}/\Tid}=\Tid\,, \\
    \muB=\frac{\muid}{\Tid}\frac{{\cal A}}{{\cal A}/\Tid}=\muid \,,
\end{eqnarray}
where
\begin{align}
    \nonumber
    {\cal A} = & \frac{d}{2\pi^2} \int_0^\infty dk\,k^2 E(k) \frac{(E(k)-\muid)}{\Tid^2} \fFD(k)(1-\fFD(k)) \times \frac{d}{2\pi^2} \int_0^\infty dk \frac{k^2}{\Tid} \fFD(k)(1-\fFD(k)) \\
    & - \frac{d}{2\pi^2} \int_0^\infty dk\,k^2 \frac{(E(k)-\muid)}{\Tid^2} \fFD(k)(1-\fFD(k)) \times \frac{d}{2\pi^2} \int_0^\infty dk\,k^2 \frac{E(k)}{\Tid} \fFD(k)(1-\fFD(k))  \,.
\end{align}
As expected, we find that the physical temperature and baryon chemical potential correspond to the Lagrange multipliers in this case. 

\bibliography{biblio.bib}
\end{document}